\documentclass[aps,reprint,amsmath,amssymb,onecolumn,superscriptaddress]{revtex4-2}
\usepackage{graphicx}
\usepackage{dcolumn}
\usepackage{bm}
\usepackage{subfigure}
\usepackage{float}
\usepackage{color}
\usepackage{mathrsfs}
\usepackage{amstext}
\usepackage{booktabs}
\usepackage{siunitx}
\usepackage{lineno}

\begin{document}


\title{Scaling Enhancement in Distributed Quantum Sensing via Bidirectional Causal Routing}


\author{Binke Xia}
\affiliation{QICI Quantum Information and Computation Initiative, School of Computing and Data Science, The University of Hong Kong, Pokfulam Road, Hong Kong, China}
\author{Zhaotong Cui}
\affiliation{State Key Laboratory of Photonics and Communications, Institute for Quantum Sensing and Information Processing, School of Automation and Intelligent Sensing, Shanghai Jiao Tong University, Shanghai 200240, China}
\author{Jingzheng Huang}
\email{jzhuang1983@sjtu.edu.cn}
\affiliation{State Key Laboratory of Photonics and Communications, Institute for Quantum Sensing and Information Processing, School of Automation and Intelligent Sensing, Shanghai Jiao Tong University, Shanghai 200240, China}
\affiliation{Hefei National Laboratory, Hefei 230088, China}
\affiliation{Shanghai Research Center for Quantum Sciences, Shanghai 201315, China}
\author{Yuxiang Yang}
\email{yxyang@hku.hk}
\affiliation{QICI Quantum Information and Computation Initiative, School of Computing and Data Science, The University of Hong Kong, Pokfulam Road, Hong Kong, China}
\author{Guihua Zeng}
\email{ghzeng@sjtu.edu.cn}
\affiliation{State Key Laboratory of Photonics and Communications, Institute for Quantum Sensing and Information Processing, School of Automation and Intelligent Sensing, Shanghai Jiao Tong University, Shanghai 200240, China}
\affiliation{Hefei National Laboratory, Hefei 230088, China}
\affiliation{Shanghai Research Center for Quantum Sciences, Shanghai 201315, China}


\begin{abstract}
Sensing networks underpin applications ranging from fundamental physics to real-world engineering. Distributed quantum sensing (DQS) can improve measurement performance, but existing protocols typically require multipartite entanglement, which poses substantial challenges for scalable implementation. Here, we introduce a DQS protocol based on bidirectional causal routing in a cyclic network, where a single probe sequentially interrogates $M$ independent sensors along two opposite causal routes. By exploiting the noncommutativity between inter-sensor propagation and local sensing operations, the protocol turns propagation from a passive transport process into a source of sensing information, yielding an asymptotic $1/M^{2}$ scaling of the estimation precision without multipartite entanglement. We experimentally demonstrate the protocol for distributed beam-tilt sensing in a free-space quantum optical network comprising up to 9 sensors, achieving picoradian-level precision in estimating the average tilt angle. These results identify propagation dynamics and routing geometry as active metrological resources for scalable distributed quantum sensing.
\end{abstract}


\maketitle


\section{Introduction}
Quantum sensing has demonstrated capabilities of achieving higher sensitivity and precision beyond the classical counterparts \cite{RevModPhys.89.035002}, and has been applied to various scenarios ranging from gravitational wave detection \cite{LIGO_2011,PhysRevLett.123.231107,Abe_2021}, navigation technologies \cite{RevModPhys.87.637,Bongs_2019,s21165568}, biochemical applications \cite{doi:10.1021/acs.accounts.5b00484,Aslam_2023}, and materials science \cite{Casola_2018,Ho02012021}. Yet most demonstrations target a single parameter at a single location. By contrast, networking spatially distributed sensors is increasingly recognized as critical for the practical, real-world deployment of quantum sensing technologies.

A distributed sensing network generally consists of $M$ spatially separated sensors, each encoding an independent local parameter, while the quantity of interest is often a global function of these parameters, such as their average \cite{Rubio_2020,Zhang_2021}. When the sensors operate independently with localized probes, the precision for estimating such a global parameter typically follows the standard-quantum-limit scaling of $1/\sqrt{M}$ with the number of sensing nodes \cite{Rubio_2020,Zhang_2021}. Distributed quantum sensing (DQS) can improve upon this independent-sensor benchmark by distributing multipartite entanglement across the network \cite{PhysRevLett.120.080501,PhysRevA.97.032329,PhysRevA.97.042337,PhysRevLett.121.043604,PhysRevLett.121.130503,Gessner_2020,Rubio_2020,Fadel_2025}. Experimental demonstrations have employed Greenberger--Horne--Zeilinger (GHZ) polarization-entangled photon states for distributed phase sensing in discrete-variable (DV) platforms \cite{Liu_2021,PhysRevX.11.031009,Kim_2024,Liu:24}, as well as continuous-variable (CV) multipartite entanglement generated from squeezed light for sensing phase-space displacements \cite{Guo_2020}, radio-frequency signals \cite{PhysRevLett.124.150502}, and forces \cite{XiaYi_2023}. Despite these advances, extending entanglement-based DQS to larger networks remains technically challenging. Current experimental implementations involve no more than 6 nodes in DV systems and 4 nodes in CV systems, primarily due to the difficulty of generating large-scale entanglement \cite{Flamini_2019,PhysRevLett.121.250505,Jia_2025} and the fragility of non-classical probes \cite{Escher_2011,Demkowicz_2012}, thereby constraining practical deployment. These limitations motivate sensing architectures that avoid multipartite entanglement while accessing sources of metrological information not used in conventional DQS protocols.

Instead of the parallel sensing architecture commonly used in conventional DQS protocols, here we consider a cyclic network in which a single probe is routed sequentially through $M$ independent sensors. Sequential strategies are known to provide quantum-enhanced precision in single-parameter estimation without requiring multipartite entanglement \cite{PhysRevLett.96.010401,Giovannetti_2011,PhysRevLett.113.250801}. In a spatially distributed network, however, the probe must also propagate between neighboring nodes. Although such propagation is typically regarded merely as a passive transport process or as a detrimental source of loss, dispersion, and noise \cite{RevModPhys.77.513,RevModPhys.84.621,PhysRevD.101.042003}, we show that the noncommutativity between propagation and sensing processes can itself provide a useful metrological resource \cite{PhysRevLett.133.090801,PhysRevLett.133.190801,3jlc-lb5c}. The metrological role of noncommuting quantum processes has previously been explored in protocols based on quantum scrambling \cite{Manju2025Scrambling} and indefinite causal order (ICO) \cite{PhysRevA.88.022318,PhysRevLett.124.190503}, in which coherent control of alternative operation orders can enhance parameter estimation \cite{PhysRevLett.130.070803} and generate nonlinear responses to a global geometric phase \cite{PhysRevLett.124.190503,Yin_2023}.

To harness the noncommuting-propagation induced information for estimating a global network parameter, we employ bidirectional causal routing in a spatial sensing network. The probe interrogates the same set of sensors along two oppositely directed but individually well-defined routes. Because inter-sensor propagation and local sensing operations do not commute, the information encoded in the final probe state depends on the route through the network. Combining the two routes then isolates and enhances the contribution associated with the global parameter. Our approach differs both operationally and conceptually from existing metrological schemes based on noncommuting processes: the scaling advantage does not require a coherent superposition of the routes. Instead, a probabilistic mixture implemented through classical route control is sufficient, substantially reducing the experimental requirements relative to quantum-switch-based ICO protocols.

Fundamentally, bidirectional causal routing accesses propagation-induced information that is not extracted by conventional DQS readout strategies. The noncommutativity between sensing and propagation causes the parameter-dependent contributions to accumulate across the network, producing a response to the global average that contains a term proportional to $M^{2}$. Under the same network geometry and probe-resource constraints, conventional DQS uses propagation only to distribute or transport the probes and therefore does not access this contribution. Combining the opposite routes further suppresses correlations with nuisance parameters while retaining the accumulated response to the global parameter, allowing even a probabilistic mixture of the routes to outperform either route individually without requiring a coherent route superposition.

We implement this protocol for the distributed sensing of transverse momentum kicks, corresponding to beam tilts in a free-space quantum optical network \cite{PhysRevLett.102.173601,Xia_2023}. In this setting, the momentum-kick operation and free-space propagation do not commute, causing each local momentum kick to induce an additional position displacement during subsequent propagation. When the probe sequentially interrogates all sensors along oppositely directed causal routes, the average momentum kick is mapped onto a route-dependent displacement of the final probe state. For a fixed average inter-node distance, the corresponding response contains an $M^{2}$ contribution and yields an asymptotic precision limit scaling as $1/M^{2}$ under fixed probe resources. We experimentally demonstrate this scaling enhancement in a free-space network comprising up to 9 sensors. Weak-value amplification (WVA) is employed as a practical readout technique to convert the propagation-induced position displacement into a measurable momentum shift while suppressing technical noise \cite{PhysRevX.4.011031}. The experiment achieves picoradian-level precision in estimating the average beam-tilt angle. These results identify bidirectional causal routing and propagation-induced information as previously unexplored design resources for scalable distributed quantum sensing, with potential applications in interferometric alignment, vibration and acoustic monitoring, wavefront sensing, and beam-pointing stabilization.

\section{Results}
\subsection{Theoretical Framework}
In a general distributed quantum sensing task, $M$ individual parameters $\theta_{1}$, $\theta_{2}$, ..., $\theta_{M}$ are encoded into a quantum probe distributed over a sensing network comprising $M$ spatially separated sensors. The goal is typically not to estimate all parameters independently, but rather to infer a weighted collective parameter, most commonly the average value $\bar{\theta}=\sum_{j=1}^{M}\theta_{j}/M$, from final probe state. Thus, distributed quantum sensing can be formulated as the estimation of a single parameter of interest in the presence of nuisance parameters, rather than as a generic multiparameter estimation problem. Specifically, the original parameter vector can be reparameterized as $\bm{\xi}=(\xi_{1},\xi_{2},\dots,\xi_{M})^{\mathsf{T}}$, where $\xi_{\mathrm{I}}=\xi_{1}=\bar{\theta}$ denotes the parameter of interest, and the remaining components $\bm{\xi}_{\mathrm{N}}=(\xi_{2},\dots,\xi_{M})^{\mathsf{T}}$ span the complementary parameter subspace and serve as nuisance parameters that affect the attainable precision in estimating $\xi_{\mathrm{I}}$. Without loss of generality, one may choose $\xi_{2}=\theta_{2}$, ..., $\xi_{M}=\theta_{M}$. In this setting, neither the quantum Fisher information matrix (QFIM), $\mathcal{Q}(\bm{\theta})$ or $\mathcal{Q}(\bm{\xi})$, nor the single-parameter quantum Fisher information (QFI) associated with $\bar{\theta}$ alone directly determines the ultimate precision limit, namely the quantum Cramér--Rao bound (QCRB), for estimating the parameter of interest. Instead, the precision limit is governed by the partial QFI (equivalently, the Schur complement of the nuisance-parameter block) \cite{Suzuki_2020},
\begin{equation}
    \mathcal{Q}_{\xi_{\mathrm{I}}|\bm{\xi}_{\mathrm{N}}}(\bm{\xi}) = \mathcal{Q}_{\xi_{\mathrm{I}};\xi_{\mathrm{I}}}(\bm{\xi})-\mathcal{Q}_{\xi_{\mathrm{I}};\bm{\xi}_{\mathrm{N}}}(\bm{\xi})\left[\mathcal{Q}_{\bm{\xi}_{\mathrm{N}};\bm{\xi}_{\mathrm{N}}}(\bm{\xi})\right]^{-1}\mathcal{Q}_{\bm{\xi}_{\mathrm{N}};\xi_{\mathrm{I}}}(\bm{\xi}), \label{eq:1}
\end{equation}
whose inverse gives the achievable QCRB for estimating $\xi_{\mathrm{I}}=\bar{\theta}$. Throughout this work, we assume that all $M$ sensors are identical, i.e., each sensor is governed by the same sensing Hamiltonian, denoted by $\hat{H}_{S}$. The $j$-th sensor encodes its local unknown parameter $\theta_{j}$ onto the probe state via a unitary process $\hat{U}_{\theta_{j}}=\exp(-\mathrm{i}\hat{H}_{S}\theta_{j})$.

\begin{figure}[htb]
	\centering
	\includegraphics[width=0.75\linewidth]{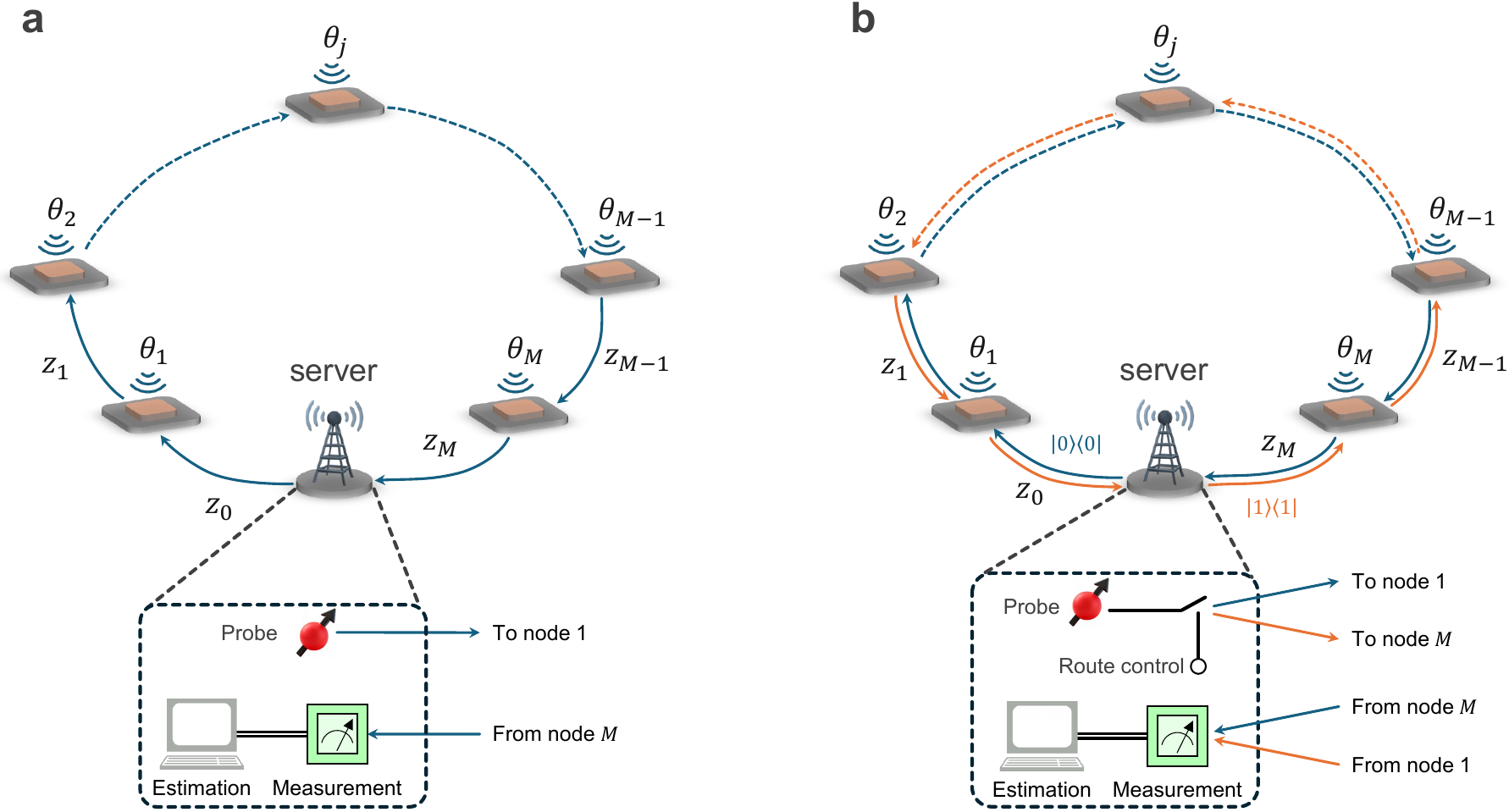}
	\caption{\label{fig:1} Schematic of distributed quantum sensing within a cyclic network. \textbf{a}, A single probe sequentially interrogates $M$ sensing nodes along a fixed causal route. \textbf{b}, Bidirectional causal routing combines the forward and reverse routes through either coherent or classical route control. The ancillary degree of freedom labels or controls the propagation direction, while probe preparation and final measurement are performed at the server node.}
\end{figure}

Existing schemes for distributed quantum sensing typically consider star-type networks, in which the probe state is routed to all sensors in parallel \cite{Zhang_2021}. Here, by contrast, we consider a cyclic network consisting of a server node and $M$ sensing nodes, as depicted in Fig.~\ref{fig:1}. An initial probe state $|\psi_{i}\rangle$ is prepared at the server node and interrogates $M$ independent sensors sequentially before returning to the server, where the final state $|\psi_{f}\rangle$ is measured to estimate the average parameter $\bar{\theta}$. For a fixed causal route and in the absence of propagation effects, this setting reduces to an ideal sequential quantum metrological protocol. The total sensing evolution is then described by $\hat{U}(\bm{\theta})=\exp[-\mathrm{i}\hat{H}_{S}\sum_{j=1}^{M}\theta_{j}]=\exp[-\mathrm{i}\hat{H}_{S}M\bar{\theta}]$, which depends exclusively on the parameter of interest. Consequently, the ultimate precision for estimating the parameter $\bar{\theta}$ scales linearly with the number $M$ of sensors, i.e., $\delta\bar{\theta}_{\min}\propto 1/M$.

In realistic sensing networks, however, one must also account for the free propagation of the probe between nodes. In this work, we investigate the propagation process that can be described via a unitary operator $\hat{U}_{z}=\exp[-\mathrm{i}\hat{H}_{P}z]$ (or $\hat{U}_{t}=\exp[-\mathrm{i}\hat{H}_{P}t]$), where $z$ (or $t$) denotes the propagation distance (or time) and $\hat{H}_{P}$ is the propagation Hamiltonian. The achievable precision can be degraded when the propagation and sensing Hamiltonians do not commute, $[\hat{H}_{P},\hat{H}_{S}]\neq 0$. In this case, the noncommutative propagation dynamics can drive the probe away from the optimal sensing trajectory and induce detrimental correlations between the parameter of interest and the nuisance parameters. By contrast, if the probe interrogates the sensors in opposite causal routes and the corresponding measurement outcomes are combined, these detrimental correlations can, in principle, be suppressed. Moreover, the noncommutativity between propagation and sensing dynamics may be harnessed as a metrological resource, enabling a further precision enhancement that can potentially surpass the conventional linear scaling.

To be specific, we consider a free-space quantum optical sensing scenario. In this context, the sensing process at each sensor is described by a unitary evolution $\hat{U}_{\theta}=\exp(-\mathrm{i}\theta\hat{X})$, which imparts a momentum displacement, or momentum kick, of magnitude $\theta$ to the probe state. Experimentally, such an operation can be implemented, for example, by introducing an angular tilt to the optical beam. The free-space propagation of the probe between nodes is described by the unitary operator $\hat{U}_{z}=\exp(-\mathrm{i}z\hat{P}^{2}/2k)$, where $k$ is the wave number and $z$ denotes the propagation distance. This sensing scenario is relevant to a variety of practical applications, such as optical path alignment, vibration sensing, and acoustic sensing.

We denote the propagation distances between consecutive nodes by $z_{0}$, $z_{1}$, $\cdots$, $z_{M}$, where $z_{0}$ is the distance from the server node to the first sensor, and $z_{M}$ is the distance from the last sensor back to the server node. We first consider the case in which the probe sequentially interrogates all sensors in a fixed causal route. The corresponding total evolution is given by, up to an irrelevant global phase (see Supplementary Information for details)
\begin{align}
    \hat{U}^{(\mathrm{FCR})} &= \hat{U}_{+} = \hat{U}_{z_{M}}\hat{U}_{\theta_{M}}\cdots\hat{U}_{\theta_{2}}\hat{U}_{z_{1}}\hat{U}_{\theta_{1}}\hat{U}_{z_{0}} \nonumber\\
    &= \exp\left[-\mathrm{i}\frac{(M+1)\bar{z}}{2k}\hat{P}^{2}\right]\exp\left(-\mathrm{i}\frac{g_{1}}{k}\hat{P}\right)\exp\left[-\mathrm{i}\frac{g_{1}+g_{2}}{(M+1)\bar{z}}\hat{X}\right], \label{eq:2}
\end{align}
where $\bar{z}=\frac{1}{M+1}\sum_{j=0}^{M}z_{j}$ is the average propagation distance between adjacent nodes, and
\begin{equation}
	g_{1} = \sum_{j=0}^{M-1}z_{j}\left(\sum_{l=j+1}^{M}\theta_{l}\right),\; g_{2} = \sum_{j=1}^{M}z_{j}\left(\sum_{l=1}^{j}\theta_{l}\right),\; g_{1}+g_{2}=(M+1)M\bar{z}\bar{\theta}. \label{eq:3}
\end{equation}
The final probe state is therefore $|\psi_{+}\rangle=\hat{U}_{+}|\psi_{i}\rangle$, where $|\psi_{i}\rangle$ is the initial probe state. Eq.~(\ref{eq:2}) shows that the final state is fully characterized by the two collective parameters $g_{1}$ and $g_{2}$. Since $g_{1}+g_{2}=(M+1)M\bar{z}\bar{\theta}$, the parameter pair $(g_{1},g_{2})^{\mathsf{T}}$ can equivalently be reparameterized as $(\bar{\theta},g_{2})^{\mathsf{T}}$. Consequently, the DQS task is reduced to estimating the global parameter $\bar{\theta}$ in the presence of a single nuisance parameter $g_{2}$.

The entries of QFIM $\mathcal{Q}^{(\mathrm{FCR})}(\bar{\theta},g_{2})=\mathcal{Q}(\bar{\theta},g_{2}||\psi_{+}\rangle)$, associated with estimating $\bar{\theta}$ and $g_{2}$ from the sequential final state $|\psi_{+}\rangle$, are
\begin{equation}
    \label{eq:4}
    \begin{split}
        \mathcal{Q}^{(\mathrm{FCR})}_{\bar{\theta};\bar{\theta}} &= \frac{4(M+1)^{2}M^{2}\bar{z}^{2}\langle\Delta\hat{P}^{2}\rangle_{i}}{k^{2}}+\frac{8(M+1)M^{2}\bar{z}\mathrm{Cov}_{i}(\hat{X},\hat{P})}{k}+4M^{2}\langle\Delta\hat{X}^{2}\rangle_{i},\\
        \mathcal{Q}^{(\mathrm{FCR})}_{\bar{\theta};g_{2}} &= \mathcal{Q}^{(\mathrm{FCR})}_{g_{2};\bar{\theta}} = -\frac{4(M+1)M\bar{z}\langle\Delta\hat{P}^{2}\rangle_{i}}{k^{2}}-\frac{4M\mathrm{Cov}_{i}(\hat{X},\hat{P})}{k},\\
        \mathcal{Q}^{(\mathrm{FCR})}_{g_{2};g_{2}} &= \frac{4\langle\Delta\hat{P}^{2}\rangle_{i}}{k^{2}},
    \end{split}
\end{equation}
where $\langle\Delta\hat{A}^{2}\rangle_{i}$ denotes the variance of the operator $\hat{A}$ and $\mathrm{Cov}_{i}(\hat{A},\hat{B})$ denotes the covariance between operators $\hat{A}$ and $\hat{B}$, both evaluated with respect to the initial probe state $|\psi_{i}\rangle$. Substituting Eq.~(\ref{eq:4}) into Eq.~(\ref{eq:1}), we obtain the partial QFI for estimating $\bar{\theta}$ in the presence of the nuisance parameter $g_{2}$,
\begin{equation}
    \mathcal{Q}^{(\mathrm{FCR})}_{\bar{\theta}|g_{2}} = 4M^{2}\left[\langle\Delta\hat{X}^{2}\rangle_{i}-\frac{\mathrm{Cov}_{i}(\hat{X},\hat{P})^{2}}{\langle\Delta\hat{P}^{2}\rangle_{i}}\right], \label{eq:5}
\end{equation}
which leads to a linear-scaling precision limit $\delta\bar{\theta}_{\min}^{(\mathrm{FCR})}\propto 1/M$. Although the QFIM element $\mathcal{Q}_{\bar{\theta};\bar{\theta}}^{(\mathrm{FCR})}$ contains an $M^{4}$ contribution generated by the noncommutativity between sensing and propagation, this propagation-induced information is inaccessible in a single fixed causal route. It is exactly cancelled in the partial QFI by the detrimental correlation between the parameter of interest and the nuisance parameter (i.e., the off-diagonal terms in the QFIM), leaving only the conventional linear scaling in the achievable estimation precision.

To eliminate this detrimental correlation, we allow the probe state to pass through the network in a combination, either coherent or incoherent, of two opposite causal routes. The evolution associated with the reverse causal route is given by $\hat{U}_{-}=\hat{U}_{z_{0}}\hat{U}_{\theta_{1}}\hat{U}_{z_{1}}\hat{U}_{\theta_{2}}\cdots\hat{U}_{\theta_{M}}\hat{U}_{z_{M}}$. As depicted in Fig.~\ref{fig:1}\textbf{\textsf{b}}, a route-control ancilla is employed to switch the causal route of the probe within the network, with the state $|0\rangle\langle 0|$ selecting the forward-routing direction and $|1\rangle\langle 1|$ selecting the backward-routing direction. The joint evolution of the probe and the ancilla under bidirectional causal routing is then described by $\hat{U}^{(\mathrm{BCR})}=\hat{U}_{+}\otimes|0\rangle\langle 0|+\hat{U}_{-}\otimes|1\rangle\langle 1|$, which remains fully determined by the two collective parameters $g_{1}$ and $g_{2}$.

In principle, the noncommutativity-induced scaling enhancement can be extracted by combining two opposite causal routes using a route-control ancilla, $\hat{\rho}_{\mathrm{con}}^{(\mathrm{BCR})}=(|0\rangle\langle 0|+|1\rangle\langle 1|)/2$. With this bidirectional-causal-routing strategy, the final joint state of the probe and the ancilla is
\begin{align}
    \hat{\rho}_{f}^{(\mathrm{BCR})} &= \hat{U}^{(\mathrm{BCR})}\left(|\psi_{i}\rangle\langle\psi_{i}|\otimes\hat{\rho}_{\mathrm{con}}\right)\left[\hat{U}^{(\mathrm{BCR})}\right]^{\dagger} \nonumber\\
    &= \frac{1}{2}|\psi_{+}\rangle\langle\psi_{+}|\otimes|0\rangle\langle 0|+\frac{1}{2}|\psi_{-}\rangle\langle\psi_{-}|\otimes|1\rangle\langle 1|, \label{eq:6}
\end{align}
where $|\psi_{-}\rangle=\hat{U}_{-}|\psi_{i}\rangle$ is the final probe state associated with the backward-routing direction. In this case, the control ancilla provides only a classical label that records the routing direction of the probe. The QFIM for estimating $\bar{\theta}$ and $g_{2}$ from the final state $\hat{\rho}_{f}^{(\mathrm{BCR})}$ is therefore given by $\mathcal{Q}^{(\mathrm{BCR})}(\bar{\theta},g_{2})=[\mathcal{Q}(\bar{\theta},g_{2}||\psi_{+}\rangle)+\mathcal{Q}(\bar{\theta},g_{2}||\psi_{-}\rangle)]/2$, with entries
\begin{equation}
    \label{eq:7}
    \begin{split}
        \mathcal{Q}^{(\mathrm{BCR})}_{\bar{\theta};\bar{\theta}} &= \frac{2(M+1)^{2}M^{2}\bar{z}^{2}\langle\Delta\hat{P}^{2}\rangle_{i}}{k^{2}}+\frac{4(M+1)M^{2}\bar{z}\mathrm{Cov}_{i}(\hat{X},\hat{P})}{k}+4M^{2}\langle\Delta\hat{X}^{2}\rangle_{i},\\
        \mathcal{Q}^{(\mathrm{BCR})}_{\bar{\theta};g_{2}} &= \mathcal{Q}^{(\mathrm{BCR})}_{g_{2};\bar{\theta}} = -\frac{2(M+1)M\bar{z}\langle\Delta\hat{P}^{2}\rangle_{i}}{k^{2}},\\
        \mathcal{Q}^{(\mathrm{BCR})}_{g_{2};g_{2}} &= \frac{4\langle\Delta\hat{P}^{2}\rangle_{i}}{k^{2}}.
    \end{split}
\end{equation}
This gives the partial QFI
\begin{equation}
    \mathcal{Q}^{(\mathrm{BCR})}_{\bar{\theta}|g_{2}} = \frac{(M+1)^{2}M^{2}\bar{z}^{2}\langle\Delta\hat{P}^{2}\rangle_{i}}{k^{2}}+\frac{4(M+1)M^{2}\bar{z}\mathrm{Cov}_{i}(\hat{X},\hat{P})}{k}+4M^{2}\langle\Delta\hat{X}^{2}\rangle_{i}. \label{eq:8}
\end{equation}
Importantly, the $M^{4}$ contribution remains accessible in the partial QFI under bidirectional causal routing. For a fixed average distance $\bar{z}$ between adjacent nodes and fixed input-probe statistics, this contribution yields the asymptotic precision scaling $\delta\bar{\theta}_{\min}^{(\mathrm{BCR})}\propto 1/M^{2}$. In contrast to a single fixed route, bidirectional causal routing combines the complementary route-dependent encodings such that the propagation-induced contribution is no longer canceled by correlations between the parameter of interest and the nuisance parameter.

To clarify the respective roles of route coherence and route information, we compare three implementations of bidirectional routing. The classically labeled strategy described above retains the route-control ancilla in the mixed state $\hat{\rho}_{\mathrm{con}}^{(\mathrm{BCR})}=(|0\rangle\langle 0|+|1\rangle\langle 1|)/2$, so that the ancilla records the route taken by the probe. Alternatively, coherent route control is obtained by preparing the ancilla in $|\phi_{\mathrm{con}}^{(\mathrm{coh})}\rangle =(|0\rangle+|1\rangle)/\sqrt{2}$. If the route label is discarded before the final measurement, the probe is instead described by the probabilistic mixture $\hat{\rho}_{f}^{(\mathrm{mix})}=(|\psi_{+}\rangle\langle\psi_{+}|+|\psi_{-}\rangle\langle\psi_{-}|)/2$. These three strategies are connected by parameter-independent dephasing and partial-trace channels. The monotonicity of the QFIM under completely positive trace-preserving maps therefore gives $\mathcal{Q}^{(\mathrm{mix})}(\bar{\theta},g_{2})\preceq\mathcal{Q}^{(\mathrm{BCR})}(\bar{\theta},g_{2})\preceq\mathcal{Q}^{(\mathrm{coh})}(\bar{\theta},g_{2})$, where $A\preceq B$ denotes the Loewner order, such that $B-A$ is positive semidefinite. For nonsingular nuisance-parameter blocks, the monotonicity of the corresponding Schur complements further yields
\begin{equation}
    \mathcal{Q}^{(\mathrm{mix})}_{\bar{\theta}|g_{2}} \le \mathcal{Q}^{(\mathrm{BCR})}_{\bar{\theta}|g_{2}} \le \mathcal{Q}^{(\mathrm{coh})}_{\bar{\theta}|g_{2}}. \label{eq:9}
\end{equation}
The explicit expressions are provided in the Supplementary Information. In particular, in the perturbative regime $g_{1}\ll 1$, $g_{2}\ll 1$, and $\bar{\theta}\ll 1$, the probabilistic mixture attains $\mathcal{Q}^{(\mathrm{mix})}_{\bar{\theta}|g_{2}}=\mathcal{Q}^{(\mathrm{BCR})}_{\bar{\theta}|g_{2}}$, while the leading term of $\mathcal{Q}^{(\mathrm{coh})}_{\bar{\theta}|g_{2}}$ also remains proportional to $M^{4}$. Route coherence can therefore increase the available information, as reflected by the hierarchy in Eq.~(\ref{eq:9}), but is not required to retain the asymptotic $1/M^{2}$ precision scaling. In particular, the enhancement does not rely on an indefinite causal order or on a quantum-switch implementation. Instead, classically labeled bidirectional routing, and even an unlabeled probabilistic mixture in the perturbative regime, are sufficient.

At first sight, the fact that an incoherent combination can outperform either individual route may appear to conflict with the convexity of quantum Fisher information. The relevant quantity here, however, is the partial QFI for estimating $\bar{\theta}$ in the presence of the nuisance parameter $g_{2}$, rather than an individual element or the full QFIM. Although mixing cannot increase the full QFIM, the partial QFI is a Schur complement and therefore depends nonlinearly on the target--nuisance correlations. As shown by Eqs.~(\ref{eq:4}) and (\ref{eq:5}), each fixed route already contains an $M^{4}$ propagation-induced contribution in $\mathcal{Q}^{(\mathrm{FCR})}_{\bar{\theta};\bar{\theta}}$, but this contribution is canceled in $\mathcal{Q}^{(\mathrm{FCR})}_{\bar{\theta}|g_{2}}$ by its correlation with $g_{2}$. Combining the two opposite routes modifies this correlation structure and prevents the same cancellation, thereby making the previously inaccessible propagation-induced information available for estimating the global parameter. The enhancement produced by route combination thus does not originate from creating additional information through mixing, but from improving its accessibility in the presence of nuisance parameters.

The metrological advantage of our protocol therefore arises from two complementary ingredients: the noncommutativity between local sensing and inter-node propagation generates an accumulated, route-dependent encoding, while bidirectional causal routing makes this information accessible by mitigating detrimental target--nuisance correlations. Neither multipartite entanglement among the sensing nodes nor coherence between the two routes is required for the enhanced scaling. The protocol is consequently compatible with coherent optical probes and provides a substantially simpler route toward scalable distributed sensing than schemes that require large-scale entangled states or coherent control of alternative operation orders.

\subsection{Practical Implementation}
\begin{figure}[htp]
	\centering
	\includegraphics[width=0.65\linewidth]{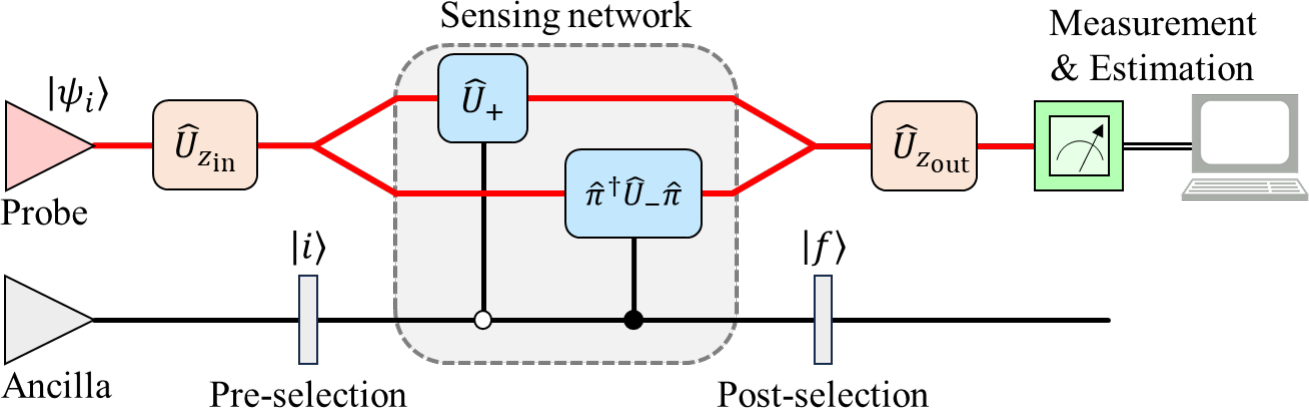}
	\caption{\label{fig:2} Schematic of the experimental implementation and weak-value readout. The polarization ancilla coherently controls the two opposite causal routes through the cyclic sensing network. A transverse parity transformation is applied to the reverse-propagating route, after which polarization postselection maps the propagation-induced position displacement onto an amplified momentum signal. The coherent route control is employed for the weak-value readout and is not required for the bidirectional-routing scaling enhancement.}
\end{figure}

To implement bidirectional causal routing experimentally, we use the transverse spatial mode of a Gaussian beam as the probe and its polarization as the route-control ancilla, as illustrated in Fig.~\ref{fig:2}. At the $j$-th sensing node, the local parameter $\theta_j$ is encoded as a transverse momentum kick, implemented by a small angular deflection of the optical beam. In practice, the probe state is initially prepared in a Gaussian profile and denoted as
\begin{equation}
	|\psi_{i}\rangle = \int\mathrm{d}x\,\psi_{i}(x)|x\rangle, \quad \psi_{i}(x) = \left(\sqrt{\frac{\pi}{2}}w_{0}\right)^{-\frac{1}{2}}\exp\left(-\frac{x^{2}}{w_{0}^{2}}\right), \label{eq:10}
\end{equation}
where $\psi_{i}(x)$ is the wave function of initial probe state and $w_{0}=2\Delta X_{i}=1/\Delta P_{i}$ is the beam-waist radius. Meanwhile, the polarization of the light beam is initialized as a $\ang{45}$ linearly polarized state $|\phi_{\mathrm{con}}\rangle=|i\rangle=(|H\rangle+|V\rangle)/\sqrt{2}$ to control the causal-route direction. Notably, although coherence between the two routes is not required to retain the enhanced scaling, coherent polarization control enables the weak-value readout adopted in our experiment. Thus, route coherence is used here as a practical resource for signal extraction rather than as the origin of the scaling enhancement in our theoretical framework.

For the weak-value readout, we introduce a transverse parity transformation on the reverse-propagating route, implemented by the reflection geometry of the free-space network. Consequently, the joint evolution of the probe and the ancilla can be represented as $\hat{U}_{+}\otimes|H\rangle\langle H|+\hat{\pi}^{\dagger}\hat{U}_{-}\hat{\pi}\otimes|V\rangle\langle V|$, where $\hat{\pi}$ denotes the parity operator, satisfying $\hat{\pi}^{\dagger}\hat{X}\hat{\pi}=-\hat{X}$ and $\hat{\pi}^{\dagger}\hat{P}\hat{\pi}=-\hat{P}$. In addition, the propagation distances $z_{\mathrm{in}}$ (from the initial probe state at the beam waist to the input of the network) and $z_{\mathrm{out}}$ (from the output of the network to the detection plane) must be taken into account. These introduce additional propagation evolutions, denoted as $\hat{U}_{z_{\mathrm{in}}}$ and $\hat{U}_{z_{\mathrm{out}}}$, acting on the probe state before entering and after exiting the sensing network, respectively. To implement the WVA technique within our experimental scheme, we ensure that the parameters $\bar{\theta}\ll 1$, $g_{1}\ll 1$ and $g_{2}\ll 1$ are maintained throughout the experiments. Under these conditions, the joint evolution of the probe and the ancilla in our scheme can be expressed as:
\begin{align}
	\hat{U}_{\mathrm{exp}} &= \hat{U}_{z_{\mathrm{out}}}\hat{U}_{+}\hat{U}_{z_{\mathrm{in}}}\otimes|H\rangle\langle H|+\hat{U}_{z_{\mathrm{out}}}\hat{\pi}^{\dagger}\hat{U}_{-}\hat{\pi}\hat{U}_{z_{\mathrm{in}}}\otimes|V\rangle\langle V| \nonumber\\
	&\approx \left(\hat{U}_{z_{\mathrm{tot}}}\otimes\hat{\mathbb{I}}\right)\left\{\hat{\mathbb{I}}-\mathrm{i}\left[\frac{\bar{z}}{2k}M^{2}\bar{\theta}+\left(\frac{\bar{z}}{2k}+\frac{z_{\mathrm{in}}}{k}\right)M\bar{\theta}\right]\hat{P}\otimes\hat{A}-\mathrm{i}\frac{g_{1}-g_{2}}{2k}\hat{P}\otimes\hat{\mathbb{I}}-\mathrm{i}M\bar{\theta}\hat{X}\otimes\hat{A}\right\}, \label{eq:11}
\end{align}
where $\hat{A}=|H\rangle\langle H|-|V\rangle\langle V|$ is a Pauli operator, $z_{\mathrm{tot}}=z_{\mathrm{in}}+(M+1)\bar{z}+z_{\mathrm{out}}$ denotes the total propagation distance of the probe. It can be seen that the sensing--propagation noncommutativity generates a route-dependent position displacement that remains linear in $\bar{\theta}$ but contains a leading contribution proportional to $M^{2}\bar{\theta}$ for fixed $\bar{z}$.

Subsequently, by post-selecting the ancilla onto a final state $|f\rangle=(\mathrm{e}^{\mathrm{i}\varepsilon}|H\rangle-\mathrm{e}^{-\mathrm{i}\varepsilon}|V\rangle)/\sqrt{2}$, we can derive the final probe state as
\begin{equation}
	|\psi_{f}\rangle = \frac{\langle f|\hat{U}_{\mathrm{exp}}|i\rangle|\psi_{i}\rangle}{\Vert\langle f|\hat{U}_{\mathrm{exp}}|i\rangle|\psi_{i}\rangle\Vert} \approx \mathcal{N}_{f}\hat{U}_{z_{\mathrm{tot}}}\left\{\hat{\mathbb{I}}-\mathrm{i}A_{w}\left[\frac{\bar{z}}{2k}M^{2}\bar{\theta}+\left(\frac{\bar{z}}{2k}+\frac{z_{\mathrm{in}}}{k}\right)M\bar{\theta}\right]\hat{P}-\mathrm{i}\frac{g_{1}-g_{2}}{2k}\hat{P}-\mathrm{i}A_{w}M\bar{\theta}\hat{X}\right\}|\psi_{i}\rangle, \label{eq:12}
\end{equation}
where $\mathcal{N}_{f}=\langle f|i\rangle/\Vert\langle f|\hat{U}_{\mathrm{exp}}|i\rangle|\psi_{i}\rangle\Vert$ is the normalization factor, $\mathcal{P}_{\mathrm{W}}=\Vert\langle f|\hat{U}_{\mathrm{exp}}|i\rangle|\psi_{i}\rangle\Vert^{2}\approx\varepsilon^{2}$ is the success probability of post-selection and
\begin{equation}
	A_{w} = \frac{\langle f|\hat{A}|i\rangle}{\langle f|i\rangle} = \mathrm{i}\cot\varepsilon \approx \mathrm{i}\frac{1}{\varepsilon} \label{eq:13}
\end{equation}
is the weak value. The imaginary weak value maps the propagation-induced position displacement, whose response contains a quadratic contribution in $M$, onto an amplified transverse-momentum shift.
As a result, the amplified signal can be extracted through a transverse momentum measurement on the final probe state. In practice, this can be implemented in an optical system by employing a Fourier lens and positioning a quadcell photodetector (QPD) at its focal plane. Accordingly, the transverse position measured at the focal plane is proportional to the input transverse momentum, with effective observable
\begin{equation}
	\hat{\Pi} = \hat{U}_{f}^{\dagger}\hat{U}_{z=f}^{\dagger}\hat{X}\hat{U}_{z=f}\hat{U}_{f} = \frac{f}{k}\hat{P}, \label{eq:14}
\end{equation}
where $\hat{U}_{f}=\exp(-\mathrm{i}k\hat{X}^{2}/2f)$ is the quadratic phase transformation imparted by the Fourier lens, $f$ is the focal length and $\hat{U}_{z=f}=\exp(-\mathrm{i}f\hat{P}^{2}/2k)$ denotes the propagation from the lens to the QPD. Therefore, this setup measures the average momentum of the final probe state, and the mean value and standard deviation of this measurement can be calculated separately as
\begin{align}
	\overline{\Pi}_{f} &= \langle\hat{\Pi}\rangle_{f} \approx \frac{2f}{k\varepsilon}\langle\Delta\hat{P}^{2}\rangle_{i}\left[\frac{\bar{z}}{2k}M^{2}+\left(\frac{\bar{z}}{2k}+\frac{z_{\mathrm{in}}}{k}\right)M\right]\bar{\theta}, \label{eq:15}\\
	\Delta \Pi_{f} &= \sqrt{\langle\Delta\hat{\Pi}^{2}\rangle_{f}} \approx \frac{f}{k}\sqrt{\langle\Delta\hat{P}^{2}\rangle_{i}}. \label{eq:16}
\end{align}
Theoretically, within $\nu$ independent identical samples, the measurement signal-to-noise ratio is given by $\mathrm{SNR}=\sqrt{\nu}\overline{\Pi}_{f}/\Delta \Pi_{f}$, where $\mathrm{SNR}=1$ yields the minimum detectable value of $\bar{\theta}$ in practice
\begin{equation}
	\delta\bar{\theta}_{\min} = \frac{k\varepsilon}{\bar{z}\sqrt{\nu\langle\Delta\hat{P}^{2}\rangle_{i}}}\cdot\frac{1}{M^{2}+\left(1+2z_{\mathrm{in}}/\bar{z}\right)M}. \label{eq:17}
\end{equation}
Therefore, our experimental scheme yields a nonlinear-enhanced-scaling precision limit $\delta\bar{\theta}_{\min}\propto 1/[M^{2}+\left(1+2z_{\mathrm{in}}/\bar{z}\right)M]$ for estimating the parameter $\bar{\theta}$. Here the scaling with $M$ is evaluated at fixed average distance $\bar{z}$ between adjacent nodes and fixed input-probe and readout resources.

\subsection{Experimental Setup and Results}
\begin{figure}[htb]
	\centering
	\includegraphics[width=0.8\linewidth]{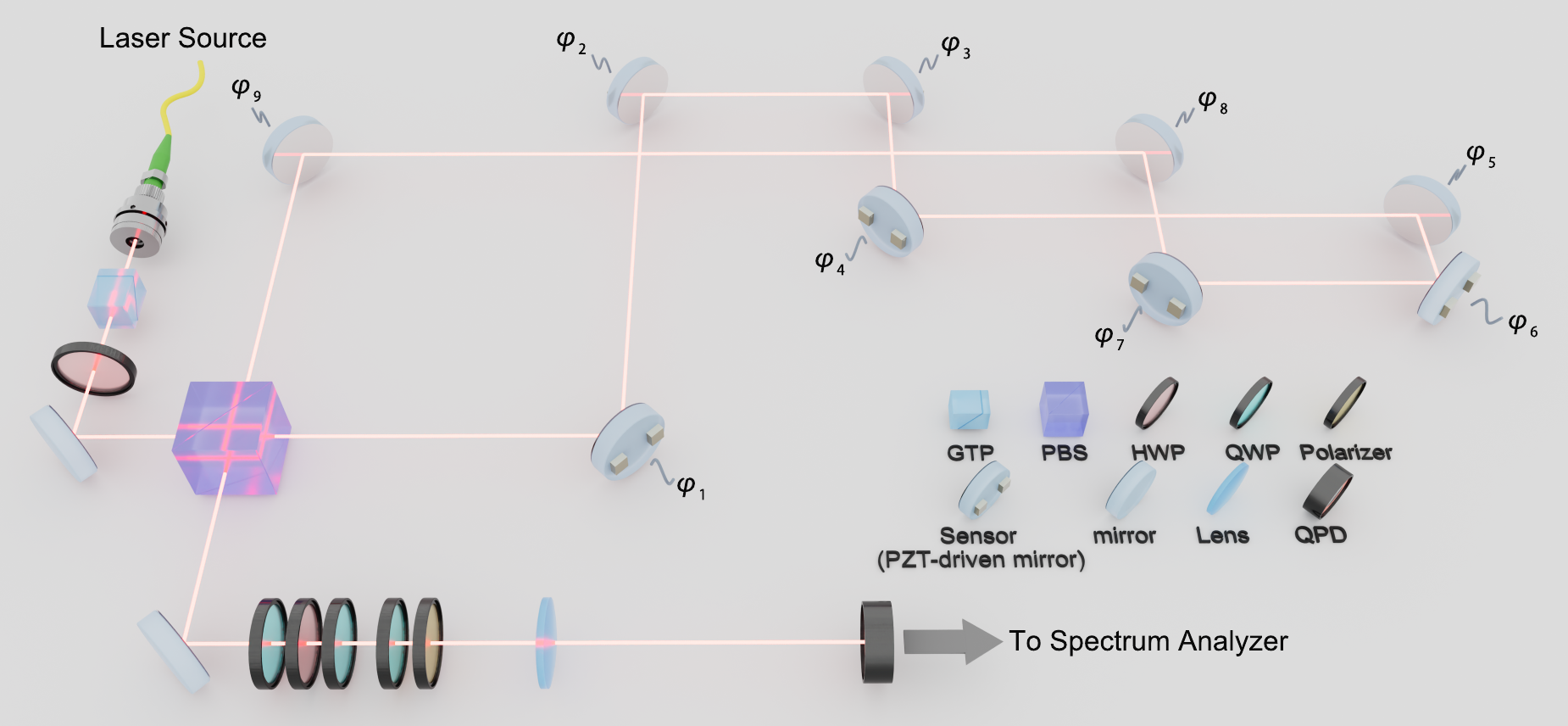}
	\caption{\label{fig:3} Experimental setup for distributed beam-tilt sensing with bidirectional causal routing. Here we illustrate a configuration with 9 sensors. A $\SI{780}{\nm}$ laser beam is coupled into free space using a collimator. The polarization is initialized through a Glan–Taylor polarizer (GTP) and a half-wave plate (HWP). The horizontal and vertical polarization components counterpropagate through the polarization Sagnac interferometer and interrogate the same set of PZT-actuated sensing mirrors in opposite directions. A QWP--HWP--QWP sequence compensates the relative polarization phase, after which polarization postselection, a Fourier lens, and a quadcell photodetector implement the weak-value momentum readout. The detector output is analyzed at the modulation frequency using a spectrum analyzer.}
\end{figure}

We realize bidirectional causal routing using a polarization Sagnac interferometer, in which the horizontally and vertically polarized components traverse the sensing nodes along opposite propagation directions. The schematic of our experimental setup is illustrated in Fig.~\ref{fig:3}, where 9 sensors are integrated, and the number of sensors in the network can be reconfigured by changing the number of loops in the Sagnac interferometer (see the Supplemental Materials for details). Mirrors driven by piezoelectric transducer (PZT) chips are utilized to generate tiny angular tilts in the light beam; specifically, a tilt angle $\varphi_{j}$ is applied at the $j$-th sensing node, resulting in a transverse momentum kick $\theta_{j} = k\varphi_{j}$ encoded onto the probe state. The interferometer is constructed using an odd number of mirrors (including both PZT-driven and non-driven mirrors), which ensures that the reverse-propagating route acquires the required transverse parity transformation. This parity operation is specific to the present weak-value readout configuration and is not required for the underlying bidirectional-routing enhancement.

The initial ancilla state $|i\rangle$ is prepared using a linear polarizer and a half-wave plate (HWP), while the post-selection state $|f\rangle$ is realized with a quarter-wave plate (QWP) and an additional linear polarizer. In our experiments, the post-selection angle $\varepsilon$ is adjusted by rotating the linear polarizer to achieve a fixed imaginary weak value $A_w\simeq-\mathrm{i}7$. Furthermore, a QWP--HWP--QWP sequence is used to compensate the unwanted relative phase accumulated between the two polarization-controlled routes. (Details on the principles and explicit configurations of these wave plates are provided in the Supplemental Materials.)

In our experiments, the differential intensity $I_{\Delta}$, which is defined as the difference between the detected intensity $I_{L}$ on the left half and $I_{R}$ on the right half of the QPD, determines the central position of the final received beam. Therefore, the measured value of $\overline{\Pi}_{f}$ is given by $I_{\Delta}$ in our experiments. Specifically, the QPD we used provides a calibrated relation $\overline{\Pi}_{f}=0.65(I_{\Delta}/I_{0})\times\text{beam radius}$ to obtain the detected beam position, where $I_{0}=I_{L}+I_{R}$ is the total intensity received by the QPD, and the received beam radius on the QPD is $2\Delta\Pi_{f}\approx(2f/k)\Delta P_{i}=2f/w_{0}k$. Thus, the directly measured signal $I_{\Delta}$ on the QPD allows us to determine the value of the sensing parameter $\bar{\theta}$ via the relation
\begin{align}
	I_{\Delta} &= \frac{2I_{0}}{1.3\varepsilon w_{0}}\left[\frac{\bar{z}}{2k}M^{2}+\left(\frac{\bar{z}}{2k}+\frac{z_{\mathrm{in}}}{k}\right)M\right]\bar{\theta} \nonumber\\
	&= \frac{\bar{z}I_{0}}{1.3\varepsilon w_{0}}\left[M^{2}+\left(1+\frac{2z_{\mathrm{in}}}{\bar{z}}\right)M\right]\bar{\varphi}, \label{eq:14}
\end{align}
where $\bar{\varphi}$ is the average value of tilt angles introduced on each sensing nodes in this network.

We stabilized the received optical power $I_{0}$ on the QPD at approximately $\SI{0.2}{\mW}$. Synchronized $\SI{10}{\kHz}$ sinusoidal drive signals from waveform generators were applied to the PZT chips to produce synchronized tilt modulation at each sensing node. A $\SI{5}{\mV}$ peak-to-peak drive corresponded to an $\SI{11}{\nano\radian}$ beam-tilt modulation (see the Methods section for details). The QPD then converted the differential optical power $I_{\Delta}$ into an electrical signal, which was fed to a spectrum analyzer to extract the $\SI{10}{\kHz}$ component corresponding to the average tilt $\bar{\varphi}$.

\begin{figure}[b]
	\centering
	\includegraphics[width=0.5\linewidth]{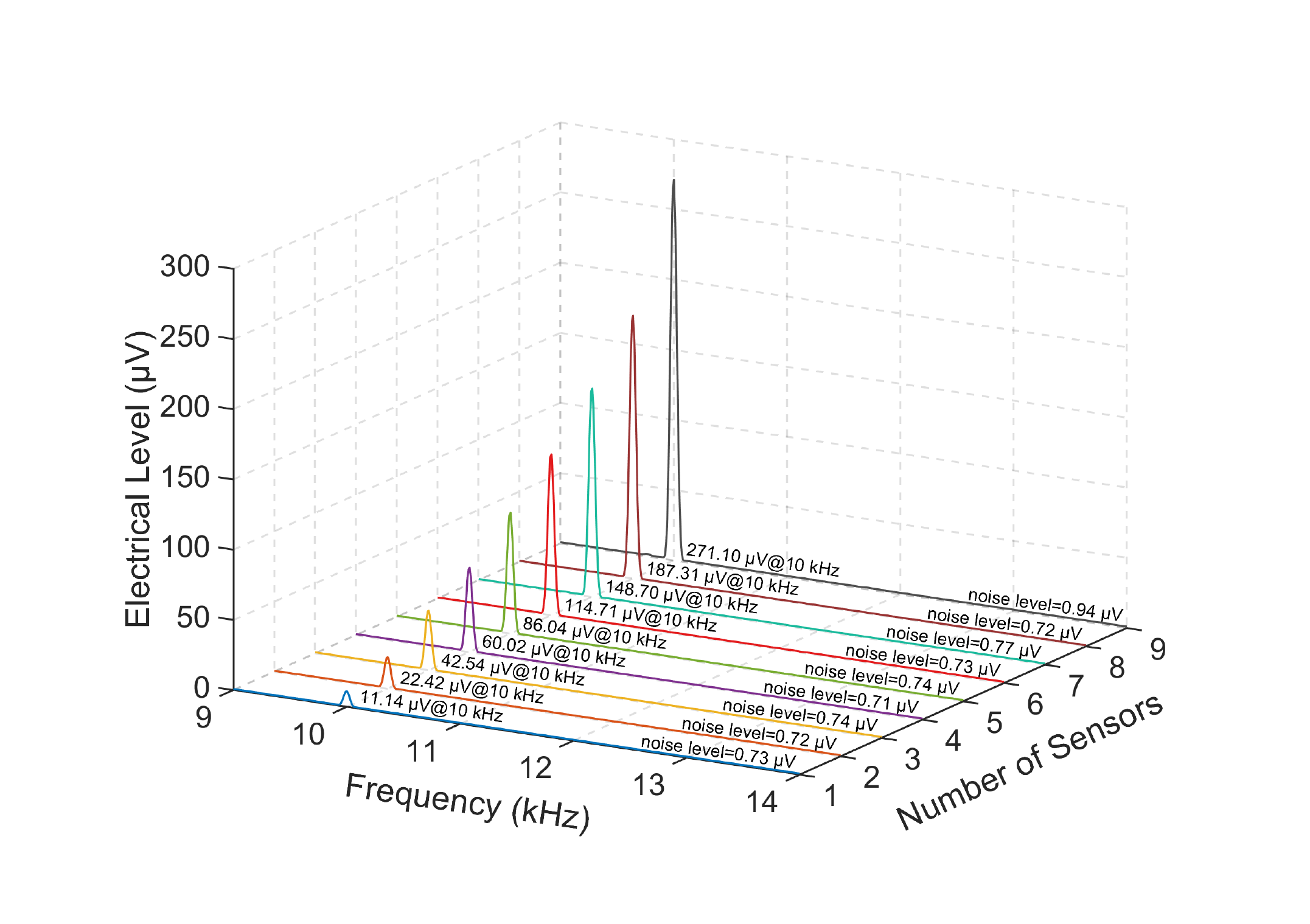}
	\caption{\label{fig:4} Detected spectrums for 1 to 9 sensors. $\SI{5}{\mV}$ peak-to-peak driving signals are synchronously applied to the PZT chips of sensors, which corresponds an $\SI{11}{\nano\radian}$ beam-tilt modulation in each sensor. The frequency of applied signals is set at $\SI{10}{\kHz}$.}
\end{figure}

Fig.~\ref{fig:4} shows the amplitude spectrums of the QPD differential signal for configurations with 1 to 9 sensors. The synchronized PZT drive amplitude was fixed at $\SI{5}{\mV}$ peak-to-peak voltage. The peak levels at $\SI{10}{\kHz}$ arise from the tilt signals $\bar{\varphi}$ and increase nonlinearly with the number of sensors. From these spectrums we also estimate the noise floor and hence the peak signal-to-noise ratio (SNR), which we use to determine the minimum detectable angular tilt $\delta\bar{\varphi}{\min}$. Because a continuous-wave coherent laser was used as the light source, the measured noise floor includes not only shot noise but also classical (technical) noises such as thermal, relative-intensity, and electronic noise. In practice, these contributions are approximately independent of the number of sensors. Accordingly, the minimum detectable tilt in our experiments satisfies (see the Methods section for a detailed analysis)
\begin{equation}
	\delta\bar{\varphi}_{\min} \propto \frac{1}{M^{2}+(1+2z_{\mathrm{in}}/\bar{z})M}, \label{eq:15}
\end{equation}
which is improved nonlinearly with the number $M$ of sensors.

\begin{figure}[htb]
	\centering
	\includegraphics[width=\linewidth]{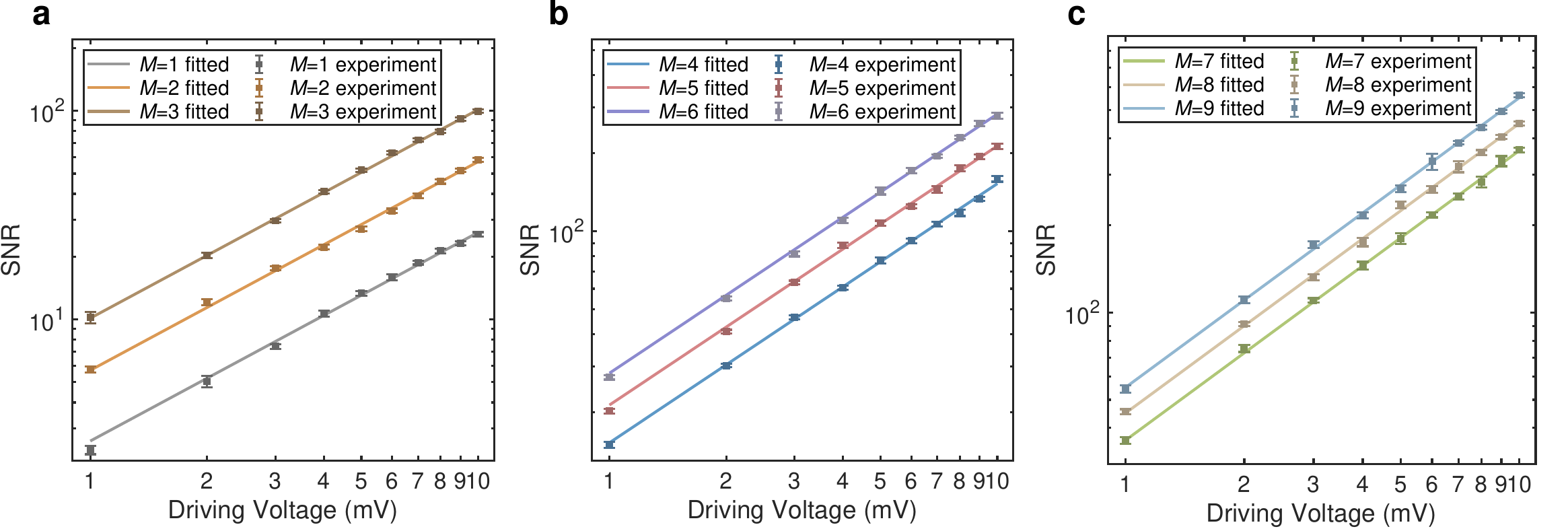}
	\caption{\label{fig:5} Experimental results of detected signal-to-noise ratio (SNR) at the spectrum analyzer. The applied peak-to-peak voltage is varied from $\SI{1}{\mV}$ to $\SI{10}{\mV}$ in $\SI{1}{\mV}$ increments, corresponding to a loaded average tilt signal from $\SI{2.2}{\nano\radian}$ to $\SI{22}{\nano\radian}$ in $\SI{2.2}{\nano\radian}$ increments. \textbf{\textsf{a}} Experimental results of 1 to 3 sensors configured in the network. \textbf{\textsf{b}} Experimental results of 4 to 6 sensors configured in the network. \textbf{\textsf{c}} Experimental results of 7 to 9 sensors configured in the network. Square markers with error bars stand for the measured SNR; solid lines represent fits to the experimental data.}
\end{figure}

To experimentally determine the minimum detectable angular tilt angle $\delta\bar{\varphi}_{\min}$ for different numbers of sensors, we synchronously varied the driving voltage applied to the PZT chips from $\SI{1}{\mV}$ to $\SI{10}{\mV}$ in $\SI{1}{\mV}$ increments, corresponding to a loaded average tilt signal from $\SI{2.2}{\nano\radian}$ to $\SI{22}{\nano\radian}$ in $\SI{2.2}{\nano\radian}$ increments. In our experiments, the number of sensors was increased from 1 to 9 in increments of 1 by reconfiguring the Sagnac interferometer. For each number of sensors and each voltage setting, we acquired 100 measurements of the spectral peak level and the noise-floor level from the spectrum analyzer, calculated the SNR for each measurement (see Supplemental Materials for details), and plotted the mean SNR with error bars versus driving voltage for each number of sensors in Fig.~\ref{fig:5}. Since the spectrum analyzer’s noise floor is independent of the tilt signal, the SNR is expected to scale linearly with the applied voltage. Accordingly, we fit a linear relation to SNR versus driving voltage for each number of sensors in Fig.~\ref{fig:5}. From the fitted lines, we can infer the driving voltage corresponding to SNR = 1 and converted it to the corresponding minimum detectable average tilt angle, $\delta\bar{\varphi}{\min}$, for each number of sensors.

\begin{figure}[htb]
	\centering
	\includegraphics[width=0.5\linewidth]{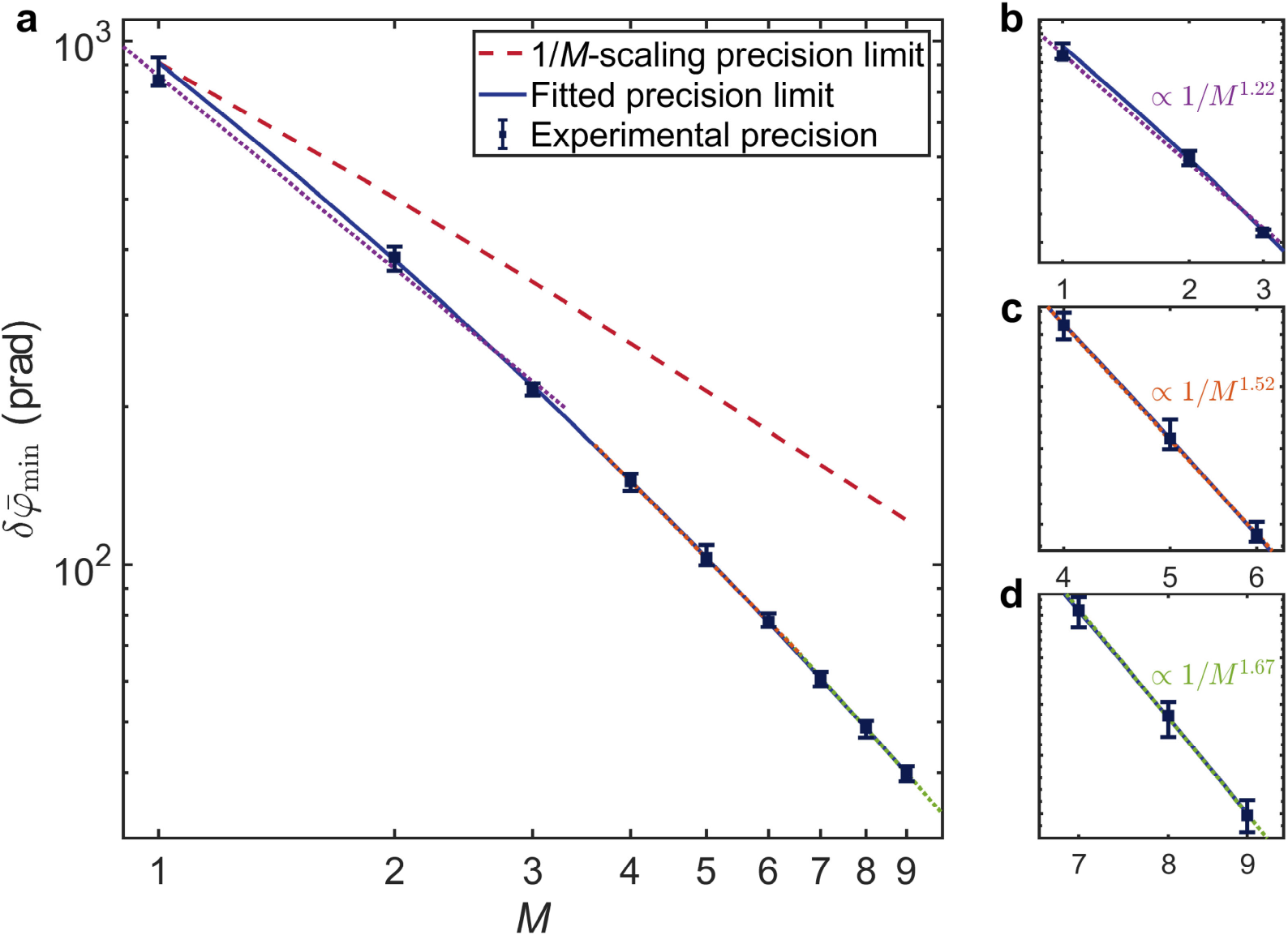}
    \caption{\label{fig:6} Experimental scaling of the measurement precision with sensor number. \textbf{\textsf{a}}, Minimum detectable average tilt angle $\delta\bar{\varphi}_{\min}$ measured for $M=1$--$9$ sensors. Square markers with error bars stand for the experimental data, the blue solid line shows the nonlinear fit $\delta\bar{\varphi}_{\min}\approx 4.77/(M^{2}+4.25M)\,\unit{\nano\radian}$, and the red dashed line denotes the corresponding $1/M$-scaling benchmark. \textbf{\textsf{b}}--\textbf{\textsf{d}}, Effective power-law fits $\delta\bar{\varphi}_{\min}\propto 1/M^{\gamma}$ for small ($M=1$--$3$), intermediate ($M=4$--$6$), and large ($M=7$--$9$) sensor-number ranges, yielding $\gamma=1.22$, $1.52$, and $1.67$, respectively. The increasing effective exponent indicates a finite-$M$ crossover from the conventional $1/M$ trend toward the asymptotic $1/M^{2}$ scaling.}
\end{figure}

In Fig.~\ref{fig:6}, we illustrate the experimental results for the minimum detectable average tilt angle, $\delta\bar{\varphi}_{\min}$, with the number of sensors varying from $M=1$ to $M=9$. These results indicate that we achieve picoradian-scale sensitivity (see the Supplemental Materials for detailed results). According to Eq.~(\ref{eq:15}), we fit the dependence of $\delta\bar{\varphi}_{\min}$ on $M$ and obtain $\delta\bar{\varphi}_{\min}\approx 4.77/(M^{2}+4.25M)\,\unit{\nano\radian}$, as shown by the fitted precision limit in Fig.~\ref{fig:6}\textbf{\textsf{a}}. The fit exhibits a high coefficient of determination, $R^{2}=99.18\%$, indicating that the experimentally observed nonlinear enhancement with increasing sensor number is well captured by Eq.~(\ref{eq:15}). For visual reference, we also plot a normalized $1/M$ trend in Fig.~\ref{fig:6}\textbf{\textsf{a}}, obtained by replacing the $M^{2}$ term in the fitted expression with a constant term.

To further reveal how the finite-$M$ experimental data approach the asymptotic $1/M^{2}$ behavior, we divide the data into three sensor-number ranges: small $(M\in\{1,2,3\})$, intermediate $(M\in\{4,5,6\})$, and large $(M\in\{7,8,9\})$. For each range, we fit the measured precision to an effective power-law dependence of the form $\delta\bar{\varphi}_{\min}\propto 1/M^{\gamma}$, with the corresponding fits shown by the dotted purple, orange, and green lines, respectively. As shown in Figs.~\ref{fig:6}\textbf{\textsf{b}}--\textbf{\textsf{d}}, the fitted effective exponents are $\gamma=1.22$, $1.52$, and $1.67$ for the small-, intermediate-, and large-$M$ ranges, respectively. These increasing exponents show that, even within the experimentally accessible finite range of sensor numbers, the scaling behavior gradually departs from the conventional $1/M$ trend and moves toward the asymptotic $1/M^{2}$ trend predicted by our theory. Therefore, the experimental results provide clear evidence of a finite-$M$ crossover toward beyond-linear scaling with respect to the number of sensors. Importantly, this enhancement is achieved without using entangled probes and remains observable in the presence of classical noise.

\section{Discussion}
Our results establish bidirectional causal routing as a mechanism for accessing propagation-induced information in distributed quantum sensing. When local sensing and inter-node propagation do not commute, each sensing interaction generates an additional route-dependent change of the probe state that accumulates as the probe traverses the network. In a single causal route, the leading $M^{4}$ contribution is already involved in the QFIM but is inaccessible due to detrimental correlations between the global parameter and nuisance parameters. Combining opposite causal routes modifies this correlation structure and makes the accumulated contribution accessible, producing a response that remains linear in the global parameter but contains a quadratic contribution in $M$. This mechanism yields an asymptotic precision scaling of $1/M^{2}$, which fundamentally arises from extracting information jointly encoded by sensing and propagation that is not accessed by conventional DQS readout strategies, rather than from multipartite entanglement or coherence between the two routes.

In our experiment, WVA is employed as a practical readout technique rather than as the origin of the enhanced scaling. The imaginary weak value maps the propagation-induced position displacement onto an amplified transverse-momentum signal, providing a convenient detector-level readout in the presence of spatial broadening and technical noise \cite{PhysRevX.4.011031}. The pure polarization ancilla, transverse parity transformation, and polarization postselection used in our implementation are required for this particular WVA readout, but not for the underlying bidirectional-routing mechanism. Without WVA, the propagation-induced information could instead be extracted through spatial-mode-resolved measurements, such as projections onto appropriate Hermite--Gaussian modes \cite{PhysRevApplied.13.034023,Xia:22}. Beyond avoiding ancilla postselection and parity control, such mode-projective measurements may facilitate the extension of our protocol to genuinely nonclassical probes, including single-photon states and quadrature-squeezed light, through mode-resolved photon counting or homodyne detection \cite{10.1063/1.4869819}.

More broadly, our results are connected to a growing class of quantum sensing and metrological schemes that exploit noncommutativity as a resource. Quantum-scrambling-based sensing uses noncommuting dynamics to reshape parameter-encoding generators and improve metrological distinguishability \cite{doi:10.1126/science.adg9500,Liu2024ButterflyMetrology,wb5z-y5y5,Manju2025Scrambling}, whereas indefinite-causal-order protocols coherently control alternative operation orders to convert noncommutativity-induced geometric phases into observable signals \cite{PhysRevLett.124.190503,Yin_2023,PRXQuantum.2.010320}. In contrast, our framework introduces a distinct mechanism, it uses the natural noncommutativity between spatial propagation and local sensing operations along two opposite but individually well-defined causal routes. Although coherent route control can increase the available information, it is not required to retain the enhanced scaling. Classically labeled bidirectional causal routing is sufficient, and even an unlabeled probabilistic mixture attains the same partial QFI in the perturbative regime. The metrological advantage therefore originates from making an existing propagation-induced encoding accessible in the presence of nuisance parameters, rather than from creating additional information through coherent control of operation orders.

Recent work has provided a general operator-algebraic framework for noncommutativity as a metrological resource \cite{3jlc-lb5c}. In distributed sensing networks, however, our results reveal a distinct mechanism: the information generated by noncommuting sensing and propagation dynamics must not only accumulate across the network but also remain accessible in the presence of nuisance parameters. Noncommutativity alone therefore does not guarantee a metrological advantage. Bidirectional causal routing provides this accessibility by modifying the target--nuisance correlation structure and retaining the accumulated propagation-induced contribution to the global parameter. This network-level mechanism distinguishes our protocol from schemes in which the enhancement is attributed primarily to the algebraic structure of noncommuting operations. It also suggests extensions to more general sensing and propagation generators, as well as to weighted linear functions of distributed parameters through node-dependent interaction times or coupling strengths.

The reported scaling assumes a fixed mean distance between adjacent nodes, so the physical extent and total propagation length of the network increase with $M$. Such growth is intrinsic to expanding a spatially distributed network, conventional DQS likewise requires additional links to connect newly added sensors, increasing the aggregate transmission resources. The key distinction is therefore not whether propagation grows with network size, but whether it is used merely for probe distribution or actively converted into sensing information through noncommuting dynamics and bidirectional causal routing. More broadly, our results identify routing geometry and propagation dynamics as active design elements for quantum sensing networks, providing a path toward scalable distributed sensing without requiring multipartite entanglement.

\section{Methods}
\subsection{Experimental materials}
In our experiments, we used a single-frequency laser (NKT Photonics Koheras HARMONIK series) operating at a wavelength of $\SI{780}{\nm}$. The probe state is initialized as Gaussian-profile beam with a beam width of $\SI{2}{\mm}$ through a collimator. A Glan–Taylor polarizer (GTP) from Thorlabs and an HWP from LBTEK are used to initialize the polarization state of light as $\ang{45}$-polarized state. We set up a polarized Sagnac interferometer to implement the cyclic sensing network with a coherent superposition of opposite causal orders, where the number of sensors in the network can be scaled by adjusting the interferometer. (See the Supplemental Materials for details.) Furthermore, the unwanted relative phase induced by the interferometer is compensated through two QWPs and a HWP from LBTEK. (See the Supplemental Materials for configurations.)

To generate an angular tilt signal on each sensing node, we pasted 2 PZT chips along the horizontal plane with an interval of $\SI{20}{\mm}$ on the back of a mirror. (See the Supplemental Materials for detailed design of sensor.) These PZT chips are from Core Tomorrow Company (Part No. NAC2013), they shift approximately $\SI{22}{\nm}$ when a voltage of $\SI{1}{\V}$ is applied. We applied two sinusoidal drive signals with opposite phases at a frequency of $f=\SI{10}{\kHz}$ to the two PZT chips on each signal mirror. These sinusoidal driving signals were generated by the wave generator of Moku:Pro from Liquid Instruments. We also employed a rubidium clock (Part No. STW-FS725) from Synchronization Technology to synchronize signals applied to PZT chips on different signal mirrors in our experiments.

To demodulate the average value of tilt angles, we first employed a QWP from LBTEK and a polarizer from Thorlabs to implement the weak value amplification with a fixed imaginary weak value $A_{w}\approx\mathrm{i}7$. Subsequently, a Fourier lens and a (QPD) (Part No. 2901) from Newport were used to measure the beam position. The received optical power on the QPD was fixed around $\SI{0.2}{\mW}$ in our experiments. The effective area of the QPD is $\qtyproduct[product-units=power]{3 x 3}{\mm}$. The QPD has a responsivity of $\SI{0.5}{\A/\W}$ at the wavelength around $\SI{780}{\nm}$. In our experimental setup, the transresistance gain was set to $\SI{20}{\kV/\A}$. Finally, the detected signal on the QPD was analyzed by the spectrum analyzer of Moku:Pro from Liquid Instruments.

\subsection{SNR calculation and experimental precision}
In our experiments, we directly measure the differential intensity using the QPD, which converts the received optical power to an electrical voltage. According to Eq.~(\ref{eq:14}), the output of QPD is a voltage signal $V_{\Delta}=\gamma I_{\Delta}$, where $\gamma$ is the conversion gain, which is the product of the responsivity and transresistance gain of the QPD. Therefore, the detected signal in our experiments satisfies $V_{\Delta}\propto [M^{2}+(1+2z_{\mathrm{in}}/\bar{z})M]\bar{\varphi}$, which contains a quadratic contribution in $M$ and linearly increases with the average tilt angle $\bar{\varphi}$. The loaded angular tilt signals in our experiments are synchronized sinusoidal signals at $\SI{10}{\kHz}$. To demodulate the average tilt angle, we then input the detected signal from the QPD to a spectrum analyzer. As shown in Fig.~\ref{fig:4}, we can then determine peak SNR as
\begin{equation*}
	\mathrm{SNR} = \frac{V_{\Delta}^{(\SI{10}{\kHz})}}{V_{\mathrm{noise}}},
\end{equation*}
where $V_{\Delta}^{(\SI{10}{\kHz})}$ corresponds to the $\SI{10}{\kHz}$ signal arising from angular tilts, and $V_{\mathrm{noise}}$ is the noise floor in the device. Principally, the detected noise floor in the spectrum analyzer consists of shot noise, electrical noise and thermal noise, which are independent of the number of sensors and the amplitude of tilt signals. Thus, the detected peak SNR in the spectrum analyzer satisfies
\begin{equation}
	\mathrm{SNR} \propto \left[M^{2}+\left(1+2\frac{z_{\mathrm{in}}}{\bar{z}}\right)M\right]\bar{\varphi}^{(\SI{10}{\kHz})}, \label{eq:16}
\end{equation}
which also contains a quadratic contribution in $M$ and linearly increases with the average tilt angle $\bar{\varphi}$.

In practice, $\mathrm{SNR}=1$ indicates the minimum detectable signal in experiments. To determine the minimum detectable tilt angle under different number of sensors in our experiments, we conducted experiments with 1 to 9 sensors by adjusting the Sagnac interferometer. For every fixed number of sensors, we change the driving voltage on PZT chips from $\SI{1}{\mV}$ to $\SI{10}{\mV}$ in $\SI{1}{\mV}$ increments. By measuring the corresponding SNR, we can fit the linear relation between the detected SNR and driving voltage for different number of sensors, as shown in Fig.~\ref{fig:5}. Then the minimum detectable tilt angle under different number of sensors can be inferred by taking $\mathrm{SNR}=1$. The detailed results of minimum detectable angles are listed in the Supplemental Materials.

\begin{acknowledgments}
	This work was supported by the National Natural Science Foundation of China via the Excellent Young Scientists Fund (Hong Kong and Macau) Project 12322516, the National Natural Science Foundation of China (No. 62471289), Quantum Science and Technology-National Science and Technology Major Project (No. 2021ZD0300703 and No. 2023200300600), Shanghai Municipal Science and Technology Major Project (Grant No. 2019SHZDZX01), Guangdong Provincial Quantum Science Strategic Initiative (Grant No. GDZX2503001), and the Hong Kong Research Grant Council (RGC) through grant 17303923 and grant 17302724.
\end{acknowledgments}

\noindent {\bf Data and Materials Availability} All data needed to evaluate the conclusions in the paper are present in the paper and the Supplementary Materials.

\noindent {\bf Competing Interests} All authors declare that they have no competing interests.

\noindent {\bf Author Contribution} B.X. designed the scheme, Y.Y., J.H. and G.Z. supervised the research project, B.X. and Y.Y. constructed the theoretical model, B.X., Z.C. and J.H. carried out the experiments, B.X. and Z.C. analyzed the experimental data, B.X. wrote the manuscript with assistance from other authors. All authors have read and approved the final version of the manuscript.

\bibliography{bibliography}

\end{document}



\title{Supplementary Materials for Scaling Enhancement in Distributed Quantum Sensing via Bidirectional Causal Routing}


\author{Binke Xia}
\affiliation{QICI Quantum Information and Computation Initiative, School of Computing and Data Science, The University of Hong Kong, Pokfulam Road, Hong Kong, China}
\author{Zhaotong Cui}
\affiliation{State Key Laboratory of Photonics and Communications, Institute for Quantum Sensing and Information Processing, School of Automation and Intelligent Sensing, Shanghai Jiao Tong University, Shanghai 200240, China}
\author{Jingzheng Huang}
\email{jzhuang1983@sjtu.edu.cn}
\affiliation{State Key Laboratory of Photonics and Communications, Institute for Quantum Sensing and Information Processing, School of Automation and Intelligent Sensing, Shanghai Jiao Tong University, Shanghai 200240, China}
\affiliation{Hefei National Laboratory, Hefei 230088, China}
\affiliation{Shanghai Research Center for Quantum Sciences, Shanghai 201315, China}
\author{Yuxiang Yang}
\email{yxyang@hku.hk}
\affiliation{QICI Quantum Information and Computation Initiative, School of Computing and Data Science, The University of Hong Kong, Pokfulam Road, Hong Kong, China}
\author{Guihua Zeng}
\email{ghzeng@sjtu.edu.cn}
\affiliation{State Key Laboratory of Photonics and Communications, Institute for Quantum Sensing and Information Processing, School of Automation and Intelligent Sensing, Shanghai Jiao Tong University, Shanghai 200240, China}
\affiliation{Hefei National Laboratory, Hefei 230088, China}
\affiliation{Shanghai Research Center for Quantum Sciences, Shanghai 201315, China}



\maketitle


\section{Quantum Precision Limits in Cyclic Sensing Network}
\subsection{QFIM and partial QFI}
In distributed quantum sensing (DQS), the central task can be formulated as single-parameter estimation in the presence of nuisance parameters. To determine the ultimate precision limits, we use the partial quantum Fisher information (QFI) as the figure of merit, which can be obtained from the corresponding quantum Fisher information matrix (QFIM). For a parameterized quantum state $\hat{\rho}(\bm{\theta})$, where $\bm{\theta}=(\theta_{1},\theta_{2},\dots,\theta_{M})^{\mathsf{T}}$ is an unknown parameter vector and $\theta_{j}$ denotes its $j$-th component, the QFIM element associated with the simultaneous estimation of $\theta_{1},\theta_{2},\dots,\theta_{M}$ from $\hat{\rho}(\bm{\theta})$ is given by
\begin{equation}
	\mathcal{Q}_{jl}(\bm{\theta}) = \frac{1}{2}\mathrm{Tr}\left(\hat{\rho}(\bm{\theta})\{\hat{L}_{j},\hat{L}_{l}\}\right), \label{eq:S1}
\end{equation}
where $\{\hat{A},\hat{B}\}=\hat{A}\hat{B}+\hat{B}\hat{A}$ denotes the anticommutator of $\hat{A}$ and $\hat{B}$, and $\hat{L}_{j}(\hat{L}_{l})$ is the symmetric logarithmic derivative (SLD) for the parameter $\theta_{j}(\theta_{l})$, which is determined by
\begin{equation}
	\frac{\partial\hat{\rho}(\bm{\theta})}{\partial \theta_{j}} = \frac{1}{2}\{\hat{\rho}(\bm{\theta}),\hat{L}_{j}\}. \label{eq:S2}
\end{equation}
We can write the spectral decomposition of the parameterized state $\hat{\rho}(\bm{\theta})$ as $\hat{\rho}(\bm{\theta})=\sum_{m}p_{m}|e_{m}\rangle\langle e_{m}|$, where $p_{m}>0$ is the eigenvalue and $|e_{m}\rangle$ is the corresponding eigenstate. Then the entry of QFIM can be calculated by
\begin{equation}
	\mathcal{Q}_{jl}(\bm{\theta}) = \sum_{m}\frac{1}{p_{m}}\left(\frac{\partial p_{m}}{\partial \theta_{j}}\right)\left(\frac{\partial p_{m}}{\partial \theta_{l}}\right)+\sum_{m}4p_{m}\mathrm{Re}\left(\frac{\partial\langle e_{m}|}{\partial \theta_{j}}\frac{\partial|e_{m}\rangle}{\partial \theta_{l}}\right)-\sum_{m,n}\frac{8p_{m}p_{n}}{p_{m}+p_{n}}\mathrm{Re}\left(\frac{\partial\langle e_{m}|}{\partial \theta_{j}}|e_{n}\rangle\langle e_{n}|\frac{\partial|e_{m}\rangle}{\partial \theta_{l}}\right). \label{eq:S3}
\end{equation}
For a pure parameterized state $\hat{\rho}(\bm{\theta})=|\psi(\bm{\theta})\rangle\langle\psi(\bm{\theta})|$, Eq. (\ref{eq:S3}) can be simplified as
\begin{equation}
	\mathcal{Q}_{jl}(\bm{\theta}) = 4\mathrm{Re}\left[\frac{\partial\langle\psi(\bm{\theta})|}{\partial \theta_{j}}\frac{\partial|\psi(\bm{\theta})\rangle}{\partial \theta_{l}}-\frac{\partial\langle\psi(\bm{\theta})|}{\partial \theta_{j}}|\psi(\bm{\theta})\rangle\langle\psi(\bm{\theta})|\frac{\partial|\psi(\bm{\theta})\rangle}{\partial \theta_{l}}\right]. \label{eq:S4}
\end{equation}

Specifically, when the parameter vector $\bm{\theta}$ is encoded into a pure quantum state through a unitary evolution $\hat{U}(\bm{\theta})$, the parameterized state is $|\psi(\bm{\theta})\rangle=\hat{U}\bm{\theta})|\psi\rangle$. Then the entry of QFIM is given by
\begin{equation}
	\mathcal{Q}_{jl}(\bm{\theta}) = 4\mathrm{Cov}_{|\psi\rangle}\left(\hat{\mathcal{H}}_{j},\hat{\mathcal{H}}_{l}\right) = \left.2\langle\{\hat{\mathcal{H}}_{j},\hat{\mathcal{H}}_{l}\}\rangle-4\langle\hat{\mathcal{H}}_{j}\rangle\langle\hat{\mathcal{H}}_{l}\rangle\right|_{|\psi\rangle}, \label{eq:S5}
\end{equation}
where $\mathrm{Cov}_{|\psi\rangle}(\hat{A},\hat{B})$ denotes the covariance between operator $\hat{A}$ and operator $\hat{B}$ on quantum state $|\psi\rangle$, i.e.
\begin{equation*}
	\mathrm{Cov}_{|\psi\rangle}\left(\hat{A},\hat{B}\right) := \frac{1}{2}\langle\psi|\{\hat{A},\hat{B}\}|\psi\rangle-\langle\psi|\hat{A}|\psi\rangle\langle\psi|\hat{B}|\psi\rangle.
\end{equation*}
In Eq. (\ref{eq:S5}), $\hat{\mathcal{H}}_{j}$ and $\hat{\mathcal{H}}_{l}$ represent the generators of parameters $g_{j}$ and $g_{l}$, which can be calculated by
\begin{equation}
	\hat{\mathcal{H}}_{j} = \mathrm{i}\hat{U}^{\dagger}(\bm{\theta})\left[\frac{\partial\hat{U}(\bm{\theta})}{\partial \theta_{j}}\right]. \label{eq:S6}
\end{equation}

Typically, the goal in DQS is to estimate a scalar function $f(\bm{\theta})$ of the original parameters $\theta_{1},\theta_{2},\dots,\theta_{M}$. In this case, we reparameterize the original parameter vector as $\bm{\xi}=(\xi_{1},\xi_{2},\dots,\xi_{M})^{\mathsf{T}}$, where $\xi_{1}=f(\bm{\theta})$ is the parameter of interest, while the remaining components $\xi_{2},\xi_{3},\dots,\xi_{M}$ span the complementary parameter subspace and act as nuisance parameters. One may choose $\xi_{2}=\theta_{2}$, $\xi_{3}=\theta_{3}$, ..., $\xi_{M}=\theta_{M}$, which is an invertible reparameterization. The QFIM for simultaneously estimating $\xi_{1},\xi_{2},\dots,\xi_{M}$ is then given by
\begin{equation}
    \mathcal{Q}(\bm{\xi}) = J_{\bm{\theta}(\bm{\xi})}^{\mathsf{T}}\mathcal{Q}(\bm{\theta})J_{\bm{\theta}(\bm{\xi})}, \label{eq:S7}
\end{equation}
where $J_{\bm{\theta}(\bm{\xi})}$ is the Jacobian matrix satisfying $J_{\bm{\theta}(\bm{\xi})}=J_{\bm{\xi}(\bm{\theta})}^{-1}$, with $[J_{\bm{\xi}(\bm{\theta})}]_{jl}=\partial\xi_{j}/\partial\theta_{l}$. Subsequently, the partial QFI for estimating $\xi_{1}=f(\bm{\theta})$ in the presence of the nuisance parameters $\xi_{2},\xi_{3},\dots,\xi_{M}$ is given by
\begin{equation}
    \mathcal{Q}_{\xi_{\mathrm{I}}|\bm{\xi}_{\mathrm{N}}}(\bm{\xi}) = \mathcal{Q}_{\xi_{\mathrm{I}};\xi_{\mathrm{I}}}(\bm{\xi})-\mathcal{Q}_{\xi_{\mathrm{I}};\bm{\xi}_{\mathrm{N}}}(\bm{\xi})\left[\mathcal{Q}_{\bm{\xi}_{\mathrm{N}};\bm{\xi}_{\mathrm{N}}}(\bm{\xi})\right]^{-1}\mathcal{Q}_{\bm{\xi}_{\mathrm{N}};\xi_{\mathrm{I}}}(\bm{\xi}), \label{eq:S8}
\end{equation}
where $\xi_{\mathrm{I}}=\xi_{1}$ denotes the parameter of interest and $\bm{\xi}_{\mathrm{N}}=(\xi_{2},\xi_{3},\dots,\xi_{M})^\mathsf{T}$ denotes the nuisance-parameter vector. The inverse of $\mathcal{Q}_{\xi_{\mathrm{I}}|\bm{\xi}_{\mathrm{N}}}(\bm{\xi})$ sets the achievable quantum Cramér-Rao bound (QCRB) for estimating $\xi_{\mathrm{I}}=\xi_{1}=f(\bm{\theta})$. Specifically, the variance for estimating the target scalar function $f(\bm{\theta})$ satisfies
\begin{equation}
	\mathrm{Var}\left[f(\bm{\theta})\right] = \mathrm{Var}\left(\xi_{1}\right) \ge \frac{1}{\nu\mathcal{Q}_{\xi_{\mathrm{I}}|\bm{\xi}_{\mathrm{N}}}(\bm{\xi})}, \label{eq:S9}
\end{equation}
where $\nu$ denotes the number of independent trials. Thus, the quantum precision limit on estimating $f(\bm{\theta})$ is denoted as
\begin{equation}
	\delta f_{\mathrm{QL}} = \frac{1}{\sqrt{\nu\mathcal{Q}_{\xi_{\mathrm{I}}|\bm{\xi}_{\mathrm{N}}}(\bm{\xi})}}. \label{eq:S10}
\end{equation}

\subsection{Quantum precision limit for fixed causal route} \label{Sec:S1.2}
In our scheme, we employ a cyclic sensing network, where a single probe queries $M$ sensors sequentially. Specifically, we investigate a free-space quantum sensing scenario, where the parameter-encoding process at each sensor is described by $\hat{U}_{\theta}=\exp(-\mathrm{i}\theta\hat{X})$ and the free-space propagation between sensors is described by $\hat{U}_{z}=\exp(-\mathrm{i}z\hat{P}^{2}/2k)$. By routing a single probe sequentially through the $M$ sensors along a single fixed causal route, the total evolution through the cyclic network is
\begin{equation*}
	\hat{U}^{(\mathrm{FCR})} = \hat{U}_{+} = \hat{U}_{z_{M}}\hat{U}_{\theta_{M}}\cdots\hat{U}_{\theta_{2}}\hat{U}_{z_{1}}\hat{U}_{\theta_{1}}\hat{U}_{z_{0}}.
\end{equation*}
Considering the relations
\begin{equation}
	\label{eq:S11}
	\begin{split}
		\hat{U}_{z}^{\dagger}\hat{X}\hat{U}_{z} = \hat{X}+\frac{z}{k}\hat{P}, \quad \hat{U}_{z}^{\dagger}\hat{P}\hat{U}_{z} = \hat{P}, \\
		\hat{U}_{z}\hat{X}\hat{U}_{z}^{\dagger} = \hat{X}-\frac{z}{k}\hat{P}, \quad \hat{U}_{z}\hat{P}\hat{U}_{z}^{\dagger} = \hat{P},
	\end{split}
\end{equation}
and using the Baker–Campbell–Hausdorff (BCH) formula
\begin{equation*}
	\mathrm{e}^{\hat{A}}\mathrm{e}^{\hat{B}} = \mathrm{e}^{\hat{A}+\hat{B}+\frac{1}{2}[\hat{A},\hat{B}]}, 
\end{equation*}
if $\hat{A}$ and $\hat{B}$ commute with their commutator, i.e., $[\hat{A},[\hat{A},\hat{B}]]=[\hat{B},[\hat{A},\hat{B}]]=0$. We can then derive $\hat{U}^{(\mathrm{FCR})}$ as
\begin{align*}
	\hat{U}^{(\mathrm{FCR})} =& \exp\left[-\frac{\mathrm{i}}{2k}\sum_{j=0}^{M-1}z_{j}\left(\sum_{l=j+1}^{M}\theta_{l}\right)^{2}\right]\exp\left[-\mathrm{i}\frac{(M+1)\bar{z}}{2k}\hat{P}^{2}\right] \\
	&\quad\times\exp\left[-\frac{\mathrm{i}}{k}\sum_{j=0}^{M-1}z_{j}\left(\sum_{l=j+1}^{M}\theta_{l}\right)\hat{P}\right]\exp\left[-\mathrm{i}M\bar{\theta}\hat{X}\right] \\
	=& \exp\left[-\mathrm{i}\frac{(M+1)\bar{z}}{2k}\hat{P}^{2}\right]\exp\left[-\frac{\mathrm{i}}{k}\sum_{j=0}^{M-1}z_{j}\left(\sum_{l=j+1}^{M}\theta_{l}\right)\hat{P}\right]\exp\left[-\mathrm{i}M\bar{\theta}\hat{X}\right],
\end{align*}
where $\bar{\theta}=\frac{1}{M}\sum_{j=1}^{M}\theta_{j}$ denotes the average of the sensing parameters, and $\bar{z}=\frac{1}{M+1}\sum_{j=0}^{M}z_{j}$ denotes the average propagation distance between adjacent nodes. The first factor is a global phase for a single route and is therefore omitted from the final probe evolution. Its route-dependent contribution is retained below when coherent route control is considered. For simplicity, we denote
\begin{equation*}
	g_{1} = \sum_{j=0}^{M-1}z_{j}\left(\sum_{l=j+1}^{M}\theta_{l}\right),\; g_{2} = \sum_{j=1}^{M}z_{j}\left(\sum_{l=1}^{j}\theta_{l}\right),
\end{equation*}
thereby
\begin{equation*}
	g_{1}+g_{2}=(M+1)M\bar{z}\bar{\theta}.
\end{equation*}
Then the $\hat{U}^{(\mathrm{FCR})}$ can be denoted as
\begin{equation}
	\hat{U}^{(\mathrm{FCR})} = \exp\left[-\mathrm{i}\frac{(M+1)\bar{z}}{2k}\hat{P}^{2}\right]\exp\left(-\mathrm{i}\frac{g_{1}}{k}\hat{P}\right)\exp\left[-\mathrm{i}\frac{g_{1}+g_{2}}{(M+1)\bar{z}}\hat{X}\right], \label{eq:S12}
\end{equation}

Substituting Eq. (\ref{eq:S12}) into Eq. (\ref{eq:S6}), we can derive the generators of parameters $g_{1}$ and $g_{2}$ as
\begin{align}
	\hat{\mathcal{H}}_{1}^{(\mathrm{FCR})} &= \mathrm{i}\left[\hat{U}^{(\mathrm{FCR})}\right]^{\dagger}\left[\frac{\partial\hat{U}^{(\mathrm{FCR})}}{\partial g_{1}}\right] = \frac{\hat{X}}{(M+1)\bar{z}}+\frac{\hat{P}}{k}, \label{eq:S13}\\
	\hat{\mathcal{H}}_{2}^{(\mathrm{FCR})} &= \mathrm{i}\left[\hat{U}^{(\mathrm{FCR})}\right]^{\dagger}\left[\frac{\partial\hat{U}^{(\mathrm{FCR})}}{\partial g_{2}}\right] = \frac{\hat{X}}{(M+1)\bar{z}}. \label{eq:S14}
\end{align}
Then the QFIM of estimating parameters $g_{1}$ and $g_{2}$ is derived as
\begin{equation}
	\mathcal{Q}^{(\mathrm{FCR})}(g_{1},g_{2}) = 4\left(
	\begin{array}{cc}
		\frac{\langle\Delta\hat{X}^{2}\rangle_{i}}{(M+1)^{2}\bar{z}^{2}}+\frac{\langle\Delta\hat{P}^{2}\rangle_{i}}{k^{2}}+\frac{2\mathrm{Cov}_{i}(\hat{X},\hat{P})}{(M+1)\bar{z}k} & \frac{\langle\Delta\hat{X}^{2}\rangle_{i}}{(M+1)^{2}\bar{z}^{2}}+\frac{\mathrm{Cov}_{i}(\hat{X},\hat{P})}{(M+1)\bar{z}k} \\
		\frac{\langle\Delta\hat{X}^{2}\rangle_{i}}{(M+1)^{2}\bar{z}^{2}}+\frac{\mathrm{Cov}_{i}(\hat{X},\hat{P})}{(M+1)\bar{z}k} & \frac{\langle\Delta\hat{X}^{2}\rangle_{i}}{(M+1)^{2}\bar{z}^{2}}
	\end{array}\right), \label{eq:S15}
\end{equation}
where $\langle\Delta\hat{X}^{2}\rangle_{i}$ and $\langle\Delta\hat{P}^{2}\rangle_{i}$ represent variances of position operator $\hat{X}$ and momentum operator $\hat{P}$ on the initial probe state $|\psi_{i}\rangle$, respectively. $\mathrm{Cov}_{i}(\hat{X},\hat{P})$ represents the covariance between position operator $\hat{X}$ and momentum operator $\hat{P}$ on the initial probe state $|\psi_{i}\rangle$.

Since $g_{1}+g_{2}=(M+1)M\bar{z}\bar{\theta}$, the relation between the parameter pairs $(g_{1},g_{2})^{\mathsf{T}}$ and $(\bar{\theta},g_{2})^{\mathsf{T}}$ is
\begin{equation}
    \left(
    \begin{array}{c}
        g_{1} \\
        g_{2}
    \end{array}\right) = \left(
    \begin{array}{cc}
        (M+1)M\bar{z} & -1 \\
        0 & 1
    \end{array}\right)\left(
    \begin{array}{cc}
        \bar{\theta} \\
        g_{2}
    \end{array}\right) = J\left(
    \begin{array}{cc}
        \bar{\theta} \\
        g_{2}
    \end{array}\right). \label{eq:S16}
\end{equation}
According to Eq.~(\ref{eq:S7}), the QFIM in the reparameterization $(\bar{\theta},g_{2})^{\mathsf{T}}$ then follows from the Jacobian transformation
\begin{align*}
    \mathcal{Q}^{(\mathrm{FCR})}(\bar{\theta},g_{2}) &= J^{\mathsf{T}}\mathcal{Q}^{(\mathrm{FCR})}(g_{1},g_{2})J \\
    &= 4\left(
	\begin{array}{cc}
	    M^{2}\langle\Delta\hat{X}^{2}\rangle_{i}+\frac{(M+1)^{2}M^{2}\bar{z}^{2}\langle\Delta\hat{P}^{2}\rangle_{i}}{k^{2}}+\frac{2(M+1)M^{2}\bar{z}\mathrm{Cov}_{i}(\hat{X},\hat{P})}{k} & -\frac{(M+1)M\bar{z}\langle\Delta\hat{P}^{2}\rangle_{i}}{k^{2}}-\frac{M\mathrm{Cov}_{i}(\hat{X},\hat{P})}{k} \\
	    -\frac{(M+1)M\bar{z}\langle\Delta\hat{P}^{2}\rangle_{i}}{k^{2}}-\frac{M\mathrm{Cov}_{i}(\hat{X},\hat{P})}{k} & \frac{\langle\Delta\hat{P}^{2}\rangle_{i}}{k^{2}}
	\end{array}\right).
\end{align*}
Substituting it into Eq. (\ref{eq:S8}), we can calculate that the corresponding partial QFI for estimating $\bar{\theta}$ with nuisance parameter $g_{2}$ is
\begin{align}
    \mathcal{Q}_{\bar{\theta}|g_{2}}^{(\mathrm{FCR})} &= \mathcal{Q}_{\bar{\theta};\bar{\theta}}^{(\mathrm{FCR})}-\mathcal{Q}_{\bar{\theta};g_{2}}^{(\mathrm{FCR})}\left[\mathcal{Q}_{g_{2};g_{2}}^{(\mathrm{FCR})}\right]^{-1}\mathcal{Q}_{g_{2};\bar{\theta}}^{(\mathrm{FCR})} \nonumber\\
    &= 4M^{2}\left[\langle\Delta\hat{X}^{2}\rangle_{i}-\mathrm{Cov}_{i}(\hat{X},\hat{P})^{2}/\langle\Delta\hat{P}^{2}\rangle_{i}\right], \label{eq:S17}
\end{align}
which yields a linear-scaling precision limit with respect to the number $M$ of sensors
\begin{equation}
	\delta\bar{\theta}_{\mathrm{QL}}^{(\mathrm{FCR})} = \frac{1}{\sqrt{\nu\mathcal{Q}_{\bar{\theta}|g_{2}}^{(\mathrm{FCR})}}} = \frac{1}{2M\sqrt{\nu\left[\langle\Delta\hat{X}^{2}\rangle_{i}-\mathrm{Cov}_{i}(\hat{X},\hat{P})^{2}/\langle\Delta\hat{P}^{2}\rangle_{i}\right]}} \propto\frac{1}{M}. \label{eq:S18}
\end{equation}
Although the QFIM element $\mathcal{Q}^{(\mathrm{FCR})}_{\bar{\theta};\bar{\theta}}$ contains a leading $M^{4}$ contribution generated jointly by sensing and propagation, this contribution is canceled in the partial QFI by the correlation between $\bar{\theta}$ and the nuisance parameter $g_{2}$. The resulting estimation precision therefore retains the linear scaling $\delta\bar{\theta}_{\mathrm{QL}}^{(\mathrm{FCR})}\propto 1/M$.

\subsection{Quantum precision limit for coherently controlled bidirectional routing}
For completeness, we first analyze coherent control of the two fixed causal routes. Specifically, the route-control ancilla initialized in a pure state $|\phi_{\mathrm{con}}^{(\mathrm{coh})}\rangle=(|0\rangle+|1\rangle)/\sqrt{2}$, allowing the probe to pass through the network in a coherent superposition of two oppositely directed routes. The evolution corresponding to the inverse directed routes is $\hat{U}_{-}=\hat{U}_{z_{0}}\hat{U}_{\theta_{1}}\hat{U}_{z_{1}}\hat{U}_{\theta_{2}}\cdots\hat{U}_{\theta_{M}}\hat{U}_{z_{M}}$. Then the joint evolution of the probe and the ancilla in the network under a bidirectional causal routing is given by
\begin{equation*}
	\hat{U}^{(\mathrm{BCR})} = \hat{U}_{+}\otimes|0\rangle\langle 0|+\hat{U}_{-}\otimes|1\rangle\langle 1|,
\end{equation*}
where the evolution $\hat{U}_{-}$ can be calculated as
\begin{align*}
	\hat{U}_{-} =& \exp\left[-\frac{\mathrm{i}}{2k}\sum_{j=1}^{M}z_{j}\left(\sum_{l=1}^{j}\theta_{l}\right)^{2}\right]\exp\left[-\mathrm{i}\frac{(M+1)\bar{z}}{2k}\hat{P}^{2}\right] \\
	&\quad\times\exp\left[-\frac{\mathrm{i}}{k}\sum_{j=1}^{M}z_{j}\left(\sum_{l=1}^{j}\theta_{l}\right)\hat{P}\right]\exp\left[-\mathrm{i}M\bar{\theta}\hat{X}\right].
\end{align*}
Combining with the expression of $\hat{U}_{+}=\hat{U}^{(\mathrm{FCR})}$ and substituting $g_{1}$ and $g_{2}$, $\hat{U}^{(\mathrm{BCR})}$ is then derived as
\begin{align*}
	\hat{U}^{(\mathrm{BCR})} =& \exp\left(-\frac{\mathrm{i}}{2k}\chi_{1}\right)\exp\left[-\mathrm{i}\frac{(M+1)\bar{z}}{2k}\hat{P}^{2}\right]\exp\left(-\mathrm{i}\frac{g_{1}}{k}\hat{P}\right)\exp\left[-\mathrm{i}\frac{g_{1}+g_{2}}{(M+1)\bar{z}}\hat{X}\right]\otimes|0\rangle\langle 0| \\
	&\quad +\exp\left(-\frac{\mathrm{i}}{2k}\chi_{2}\right)\exp\left[-\mathrm{i}\frac{(M+1)\bar{z}}{2k}\hat{P}^{2}\right]\exp\left(-\mathrm{i}\frac{g_{2}}{k}\hat{P}\right)\exp\left[-\mathrm{i}\frac{g_{1}+g_{2}}{(M+1)\bar{z}}\hat{X}\right]\otimes|1\rangle\langle 1|,
\end{align*}
where
\begin{equation*}
	\chi_{1} = \sum_{j=0}^{M-1}z_{j}\left(\sum_{l=j+1}^{M}\theta_{l}\right)^{2}, \; \chi_{2} = \sum_{j=1}^{M}z_{j}\left(\sum_{l=1}^{j}\theta_{l}\right)^{2}.
\end{equation*}
Furthermore, we note that
\begin{align*}
	\chi_{1}-\chi_{2} &= (z_{0}-z_{M})\left(\sum_{l=1}^{M}\theta_{l}\right)^{2}+\sum_{j=1}^{M-1}z_{j}\left[\left(\sum_{l=j+1}^{M}\theta_{l}\right)^{2}-\left(\sum_{l=1}^{j}\theta_{l}\right)^{2}\right] \\
	&= (z_{0}-z_{M})\left(\sum_{l=1}^{M}\theta_{l}\right)^{2}+\left[\sum_{j=1}^{M-1}z_{j}\left(\sum_{l=j+1}^{M}\theta_{l}-\sum_{l=1}^{j}\theta_{l}\right)\right]\cdot\sum_{l=1}^{M}\theta_{l} \\
	&=\left[\sum_{j=0}^{M-1}z_{j}\left(\sum_{l=j+1}^{M}\theta_{l}\right)-\sum_{j=1}^{M}z_{j}\left(\sum_{l=1}^{j}\theta_{l}\right)\right]\cdot\sum_{l=1}^{M}\theta_{l} \\
	&= (g_{1}-g_{2})\cdot M\bar{\theta} = \frac{g_{1}^{2}-g_{2}^{2}}{(M+1)\bar{z}}.
\end{align*}
Ignoring a global phase of $-(\chi_{1}+\chi_{2})/4k$, the joint evolution $\hat{U}^{(\mathrm{BCR})}$ is finally given by
\begin{align}
	\hat{U}^{(\mathrm{BCR})} =& \exp\left[-\mathrm{i}\frac{g_{1}^{2}-g_{2}^{2}}{4k(M+1)\bar{z}}\right]\exp\left[-\mathrm{i}\frac{(M+1)\bar{z}}{2k}\hat{P}^{2}\right]\exp\left(-\mathrm{i}\frac{g_{1}}{k}\hat{P}\right)\exp\left[-\mathrm{i}\frac{g_{1}+g_{2}}{(M+1)\bar{z}}\hat{X}\right]\otimes|0\rangle\langle 0| \nonumber\\
	&\quad +\exp\left[\mathrm{i}\frac{g_{1}^{2}-g_{2}^{2}}{4k(M+1)\bar{z}}\right]\exp\left[-\mathrm{i}\frac{(M+1)\bar{z}}{2k}\hat{P}^{2}\right]\exp\left(-\mathrm{i}\frac{g_{2}}{k}\hat{P}\right)\exp\left[-\mathrm{i}\frac{g_{1}+g_{2}}{(M+1)\bar{z}}\hat{X}\right]\otimes|1\rangle\langle 1|. \label{eq:S19}
\end{align}

Substituting Eq. (\ref{eq:S19}) into Eq. (\ref{eq:S6}), we can derive the generators of parameters $g_{1}$ and $g_{2}$ as
\begin{align}
	\hat{\mathcal{H}}_{1}^{(\mathrm{coh})} &= \mathrm{i}\left[\hat{U}^{(\mathrm{BCR})}\right]^{\dagger}\left[\frac{\partial\hat{U}^{(\mathrm{BCR})}}{\partial g_{1}}\right] = \frac{\hat{X}}{(M+1)\bar{z}}\otimes\hat{\mathbb{I}} + \left[\frac{\hat{P}}{k}-\frac{g_{2}}{k(M+1)\bar{z}}\right]\otimes|0\rangle\langle 0|, \label{eq:S20}\\
	\hat{\mathcal{H}}_{2}^{(\mathrm{coh})} &= \mathrm{i}\left[\hat{U}^{(\mathrm{BCR})}\right]^{\dagger}\left[\frac{\partial\hat{U}^{(\mathrm{BCR})}}{\partial g_{2}}\right] = \frac{\hat{X}}{(M+1)\bar{z}}\otimes\hat{\mathbb{I}} + \left[\frac{\hat{P}}{k}-\frac{g_{1}}{k(M+1)\bar{z}}\right]\otimes|1\rangle\langle 1|. \label{eq:S21}
\end{align}
The joint initial state of the probe and the ancilla is denoted as $|\Psi_{i}\rangle=|\psi_{i}\rangle\otimes|\phi_{\mathrm{con}}^{(\mathrm{coh})}\rangle$, where $|\phi_{\mathrm{con}}^{(\mathrm{coh})}\rangle=(|0\rangle+|1\rangle)/\sqrt{2}$. Then we can calculate that
\begin{equation}
	\label{eq:S22}
	\begin{split}
		&\mathrm{Var}_{|\Psi_{i}\rangle}\left[\hat{\mathcal{H}}_{1}^{(\mathrm{coh})}\right] = \frac{\langle\Delta\hat{X}^{2}\rangle_{i}}{(M+1)^{2}\bar{z}^{2}}+\frac{\langle\Delta\hat{P}^{2}\rangle_{i}}{2k^{2}}+\left[\frac{\langle\hat{P}\rangle_{i}}{2k}-\frac{g_{2}}{2k(M+1)\bar{z}}\right]^{2}+\frac{\mathrm{Cov}_{i}(\hat{X},\hat{P})}{(M+1)\bar{z}k}, \\
		&\mathrm{Cov}_{|\Psi_{i}\rangle}\left[\hat{\mathcal{H}}_{1}^{(\mathrm{coh})},\hat{\mathcal{H}}_{2}^{(\mathrm{coh})}\right] = \frac{\langle\Delta\hat{X}^{2}\rangle_{i}}{(M+1)^{2}\bar{z}^{2}}-\left[\frac{\langle\hat{P}\rangle_{i}}{2k}-\frac{g_{2}}{2k(M+1)\bar{z}}\right]\cdot\left[\frac{\langle\hat{P}\rangle_{i}}{2k}-\frac{g_{1}}{2k(M+1)\bar{z}}\right]
		+\frac{\mathrm{Cov}_{i}(\hat{X},\hat{P})}{(M+1)\bar{z}k}, \\
		&\mathrm{Cov}_{|\Psi_{i}\rangle}\left[\hat{\mathcal{H}}_{2}^{(\mathrm{coh})},\hat{\mathcal{H}}_{1}^{(\mathrm{coh})}\right] = \mathrm{Cov}_{|\Psi_{i}\rangle}\left[\hat{\mathcal{H}}_{1}^{(\mathrm{coh})},\hat{\mathcal{H}}_{2}^{(\mathrm{coh})}\right], \\
		&\mathrm{Var}_{|\Psi_{i}\rangle}\left[\hat{\mathcal{H}}_{2}^{(\mathrm{coh})}\right] = \frac{\langle\Delta\hat{X}^{2}\rangle_{i}}{(M+1)^{2}\bar{z}^{2}}+\frac{\langle\Delta\hat{P}^{2}\rangle_{i}}{2k^{2}}+\left[\frac{\langle\hat{P}\rangle_{i}}{2k}-\frac{g_{1}}{2k(M+1)\bar{z}}\right]^{2}+\frac{\mathrm{Cov}_{i}(\hat{X},\hat{P})}{(M+1)\bar{z}k},
	\end{split}
\end{equation}
where $\langle\Delta\hat{X}^{2}\rangle_{i}$ and $\langle\Delta\hat{P}^{2}\rangle_{i}$ represent variances of position operator $\hat{X}$ and momentum operator $\hat{P}$ on the initial probe state $|\psi_{i}\rangle$, respectively. $\langle\hat{P}\rangle_{i}$ represents the initial mean momentum of the probe state $|\psi_{i}\rangle$, and $\mathrm{Cov}_{i}(\hat{X},\hat{P})$ represents the covariance between position operator $\hat{X}$ and momentum operator $\hat{P}$ on the initial probe state $|\psi_{i}\rangle$. Then the QFIM of estimating parameters $g_{1}$ and $g_{2}$ is given by
\begin{equation}
	\mathcal{Q}^{(\mathrm{coh})}(g_{1},g_{2}) = 4\left\{
	\begin{array}{cc}
		\mathrm{Var}_{|\Psi_{i}\rangle}\left[\hat{\mathcal{H}}_{1}^{(\mathrm{coh})}\right] & \mathrm{Cov}_{|\Psi_{i}\rangle}\left[\hat{\mathcal{H}}_{1}^{(\mathrm{coh})},\hat{\mathcal{H}}_{2}^{(\mathrm{coh})}\right] \\
		\mathrm{Cov}_{|\Psi_{i}\rangle}\left[\hat{\mathcal{H}}_{2}^{(\mathrm{coh})},\hat{\mathcal{H}}_{1}^{(\mathrm{coh})}\right] & \mathrm{Var}_{|\Psi_{i}\rangle}\left[\hat{\mathcal{H}}_{2}^{(\mathrm{coh})}\right]
	\end{array}\right\}, \label{eq:S23}
\end{equation}
Combining with the Jacobian matrix in Eq. (\ref{eq:S16}), we can derive the QFIM $\mathcal{Q}^{(\mathrm{coh})}(\bar{\theta},g_{2})$ in the reparameterization $(\bar{\theta},g_{2})^{\mathsf{T}}$. Then substituting it into Eq. (\ref{eq:S8}), we can calculate the partial QFI for estimating $\bar{\theta}$ in the presence of a nuisance parameter $g_{2}$ with a coherent control as
\begin{equation}
	\mathcal{Q}_{\bar{\theta}|g_{2}}^{(\mathrm{coh})} = 4M^{2}(M+1)^{2}\bar{z}^{2}\frac{\mathrm{Var}_{|\Psi_{i}\rangle}\left[\hat{\mathcal{H}}_{1}^{(\mathrm{coh})}\right]\cdot\mathrm{Var}_{|\Psi_{i}\rangle}\left[\hat{\mathcal{H}}_{2}^{(\mathrm{coh})}\right]-\mathrm{Cov}_{|\Psi_{i}\rangle}\left[\hat{\mathcal{H}}_{1}^{(\mathrm{coh})},\hat{\mathcal{H}}_{2}^{(\mathrm{coh})}\right]^{2}}{\mathrm{Var}_{|\Psi_{i}\rangle}\left[\hat{\mathcal{H}}_{1}^{(\mathrm{coh})}\right]+\mathrm{Var}_{|\Psi_{i}\rangle}\left[\hat{\mathcal{H}}_{2}^{(\mathrm{coh})}\right]-2\mathrm{Cov}_{|\Psi_{i}\rangle}\left[\hat{\mathcal{H}}_{1}^{(\mathrm{coh})},\hat{\mathcal{H}}_{2}^{(\mathrm{coh})}\right]}. \label{eq:S24}
\end{equation}
Combining with Eq. (\ref{eq:S22}), we can determine that
\begin{align*}
	&\lim\limits_{M\to\infty}\frac{\mathrm{Var}_{|\Psi_{i}\rangle}\left[\hat{\mathcal{H}}_{1}^{(\mathrm{coh})}\right]\cdot\mathrm{Var}_{|\Psi_{i}\rangle}\left[\hat{\mathcal{H}}_{2}^{(\mathrm{coh})}\right]-\mathrm{Cov}_{|\Psi_{i}\rangle}\left[\hat{\mathcal{H}}_{1}^{(\mathrm{coh})},\hat{\mathcal{H}}_{2}^{(\mathrm{coh})}\right]^{2}}{\mathrm{Var}_{|\Psi_{i}\rangle}\left[\hat{\mathcal{H}}_{1}^{(\mathrm{coh})}\right]+\mathrm{Var}_{|\Psi_{i}\rangle}\left[\hat{\mathcal{H}}_{2}^{(\mathrm{coh})}\right]-2\mathrm{Cov}_{|\Psi_{i}\rangle}\left[\hat{\mathcal{H}}_{1}^{(\mathrm{coh})},\hat{\mathcal{H}}_{2}^{(\mathrm{coh})}\right]} \\
	&\qquad\qquad= \frac{\left(\langle\Delta\hat{P}^{2}\rangle_{i}/2k^{2}+\langle\hat{P}\rangle_{i}^{2}/4k^{2}\right)^{2}-\left(\langle\hat{P}\rangle_{i}^{2}/4k^{2}\right)^{2}}{\langle\Delta\hat{P}^{2}\rangle_{i}/k^{2}+\langle\hat{P}\rangle_{i}^{2}/k^{2}} \\
	&\qquad\qquad= \frac{\langle\Delta\hat{P}^{2}\rangle_{i}}{4k^{2}},
\end{align*}
which yields that
\begin{equation}
	\lim\limits_{M\to\infty}\mathcal{Q}_{\bar{\theta}|g_{2}}^{(\mathrm{coh})}/M^{4} = \frac{\bar{z}^{2}\langle\Delta\hat{P}^{2}\rangle_{i}}{k^{2}}. \label{eq:S25}
\end{equation}
Therefore, the coherent-control strategy achieves an asymptotic $1/M^{2}$-scaling precision limit
\begin{equation}
	\lim\limits_{M\to\infty}\delta\bar{\theta}_{\mathrm{QL}}^{(\mathrm{coh})}/M^{-2} = \lim\limits_{M\to\infty}M^{2}/\sqrt{\nu\mathcal{Q}_{\bar{\theta}|g_{2}}^{(\mathrm{coh})}} = \frac{k}{\bar{z}\sqrt{\nu\langle\Delta\hat{P}^{2}\rangle_{i}}}. \label{eq:S26}
\end{equation}

\subsection{Quantum precision limit for classically labeled bidirectional routing}
Eq. (\ref{eq:S22}) to Eq. (\ref{eq:S25}) show that coherence between the route labels is not required for retaining the leading $M^{4}$ contribution to the partial QFI. This contribution is carried by the route-dependent displacement of the probe. We therefore consider classically labeled bidirectional routing, in which the forward and reverse routes are selected with equal probability while the route label remains available at the final measurement. Therefore, we employ a classically mixed state $\hat{\rho}_{\mathrm{con}}^{(\mathrm{BCR})} = (|0\rangle\langle 0|+|1\rangle\langle 1|)/2$ to probabilistically switch the causal routes of probe state in the cyclic network. In this case, the probe passes through the cyclic network in either the forward or reverse direction with equal probability. Then the final joint state of the probe and the ancilla can be derived as
\begin{equation}
	\hat{\rho}_{f}^{(\mathrm{BCR})} = \frac{1}{2}\hat{U}_{+}|\psi_{i}\rangle\langle\psi_{i}|\hat{U}_{+}^{\dagger}\otimes|0\rangle\langle 0|+\frac{1}{2}\hat{U}_{-}|\psi_{i}\rangle\langle\psi_{i}|\hat{U}_{-}^{\dagger}\otimes|1\rangle\langle 1|, \label{eq:S27}
\end{equation}
where $\hat{U}_{+}$ and $\hat{U}_{-}$ are the evolution operators of forward and backward propagations given by
\begin{equation*}
	\begin{split}
		\hat{U}_{+} = \hat{U}_{z_{M}}\hat{U}_{\theta_{M}}\cdots\hat{U}_{\theta_{2}}\hat{U}_{z_{1}}\hat{U}_{\theta_{1}}\hat{U}_{z_{0}} = \exp\left[-\mathrm{i}\frac{(M+1)\bar{z}}{2k}\hat{P}^{2}\right]\exp\left(-\mathrm{i}\frac{g_{1}}{k}\hat{P}\right)\exp\left[-\mathrm{i}\frac{g_{1}+g_{2}}{(M+1)\bar{z}}\hat{X}\right], \\
		\hat{U}_{-} = \hat{U}_{z_{0}}\hat{U}_{\theta_{1}}\hat{U}_{z_{1}}\hat{U}_{\theta_{2}}\cdots\hat{U}_{\theta_{M}}\hat{U}_{z_{M}} = \exp\left[-\mathrm{i}\frac{(M+1)\bar{z}}{2k}\hat{P}^{2}\right]\exp\left(-\mathrm{i}\frac{g_{2}}{k}\hat{P}\right)\exp\left[-\mathrm{i}\frac{g_{1}+g_{2}}{(M+1)\bar{z}}\hat{X}\right].
	\end{split}
\end{equation*}
Here the control ancilla serves as a classical label that encodes classical correlations, enabling one to distinguish the direction of the probe state in the network. In contrast to the case of a coherent control, the final joint state obtained with a classical label is a mixed state, so the corresponding QFIM cannot be calculated using generator-based formulas. However, the final state $\hat{\rho}_{f}^{(\mathrm{BCR})}$ admits the spectral decomposition with eigenvalues $p_{+}=p_{-}=1/2$ and corresponding eigenstates $|\Psi_{+}\rangle=|\psi_{+}\rangle\otimes|0\rangle$ and $|\Psi_{-}\rangle=|\psi_{-}\rangle\otimes|1\rangle$, where $|\psi_{+}\rangle=\hat{U}_{+}|\psi_{i}\rangle$ and $|\psi_{-}\rangle=\hat{U}_{-}|\psi_{i}\rangle$. Substituting it into Eq. (\ref{eq:S3}), we can calculate that the entries of QFIM for simultaneously estimating $g_{1}$ and $g_{2}$ from $\hat{\rho}_{f}^{(\mathrm{BCR})}$ is given by
\begin{align*}
	\mathcal{Q}^{(\mathrm{BCR})}_{jl}(g_{1},g_{2}) =& 2\mathrm{Re}\left(\frac{\partial\langle\psi_{+}|}{\partial g_{j}}\frac{\partial|\psi_{+}\rangle}{\partial g_{l}}\right)-2\mathrm{Re}\left(\frac{\partial\langle\psi_{+}|}{\partial g_{j}}|\psi_{+}\rangle\langle\psi_{+}|\frac{\partial|\psi_{+}\rangle}{\partial g_{l}}\right) \\
	&+2\mathrm{Re}\left(\frac{\partial\langle\psi_{-}|}{\partial g_{j}}\frac{\partial|\psi_{-}\rangle}{\partial g_{l}}\right)-2\mathrm{Re}\left(\frac{\partial\langle\psi_{-}|}{\partial g_{j}}|\psi_{-}\rangle\langle\psi_{-}|\frac{\partial|\psi_{-}\rangle}{\partial g_{l}}\right),
\end{align*}
for $j,l=1,2$, because $p_{+}=p_{-}=1/2$ are independent of parameters and $\partial_{g_{j}}|\Psi_{+}\rangle$ is orthogonal to $|\Psi_{-}\rangle$ while $\partial_{g_{j}}|\Psi_{-}\rangle$ is orthogonal to $|\Psi_{+}\rangle$. According to Eq. (\ref{eq:S4}), the QFIM $\mathcal{Q}^{(\mathrm{BCR})}(g_{1},g_{2})$ can be calculated by
\begin{equation*}
	\mathcal{Q}^{(\mathrm{BCR})}(g_{1},g_{2}) = \frac{1}{2}\mathcal{Q}(g_{1},g_{2}||\psi_{+}\rangle)+\frac{1}{2}\mathcal{Q}(g_{1},g_{2}||\psi_{-}\rangle),
\end{equation*}
where $\mathcal{Q}(g_{1},g_{2}||\psi_{+}\rangle)$ and $\mathcal{Q}(g_{1},g_{2}||\psi_{-}\rangle)$ are the QFIMs of two pure states $|\psi_{+}\rangle$ and $|\psi_{-}\rangle$, respectively. Drawing on the calculations of $\mathcal{Q}^{(\mathrm{FCR})}(g_{1},g_{2})$, we can separately calculate the QFIMs of estimating $g_{1}$ and $g_{2}$ from $|\psi_{+}\rangle$ and $|\psi_{-}\rangle$ as
\begin{align*}
	\mathcal{Q}(g_{1},g_{2}||\psi_{+}\rangle) &= \mathcal{Q}^{(\mathrm{Seq,+})}(g_{1},g_{2}) = 4\left(
	\begin{array}{cc}
		\frac{\langle\Delta\hat{X}^{2}\rangle_{i}}{(M+1)^{2}\bar{z}^{2}}+\frac{\langle\Delta\hat{P}^{2}\rangle_{i}}{k^{2}}+\frac{2\mathrm{Cov}_{i}(\hat{X},\hat{P})}{(M+1)\bar{z}k} & \frac{\langle\Delta\hat{X}^{2}\rangle_{i}}{(M+1)^{2}\bar{z}^{2}}+\frac{\mathrm{Cov}_{i}(\hat{X},\hat{P})}{(M+1)\bar{z}k} \\
		\frac{\langle\Delta\hat{X}^{2}\rangle_{i}}{(M+1)^{2}\bar{z}^{2}}+\frac{\mathrm{Cov}_{i}(\hat{X},\hat{P})}{(M+1)\bar{z}k} & \frac{\langle\Delta\hat{X}^{2}\rangle_{i}}{(M+1)^{2}\bar{z}^{2}}
	\end{array}\right), \\
	\mathcal{Q}(g_{1},g_{2}||\psi_{-}\rangle) &= \mathcal{Q}^{(\mathrm{Seq,-})}(g_{1},g_{2}) = 4\left(
	\begin{array}{cc}
		\frac{\langle\Delta\hat{X}^{2}\rangle_{i}}{(M+1)^{2}\bar{z}^{2}} & \frac{\langle\Delta\hat{X}^{2}\rangle_{i}}{(M+1)^{2}\bar{z}^{2}}+\frac{\mathrm{Cov}_{i}(\hat{X},\hat{P})}{(M+1)\bar{z}k} \\
		\frac{\langle\Delta\hat{X}^{2}\rangle_{i}}{(M+1)^{2}\bar{z}^{2}}+\frac{\mathrm{Cov}_{i}(\hat{X},\hat{P})}{(M+1)\bar{z}k} & \frac{\langle\Delta\hat{X}^{2}\rangle_{i}}{(M+1)^{2}\bar{z}^{2}}+\frac{\langle\Delta\hat{P}^{2}\rangle_{i}}{k^{2}}+\frac{2\mathrm{Cov}_{i}(\hat{X},\hat{P})}{(M+1)\bar{z}k}
	\end{array}\right).
\end{align*}
Finally, the QFIM of estimating $g_{1}$ and $g_{2}$ using a classical label is given as
\begin{equation}
	\mathcal{Q}^{(\mathrm{BCR})}(g_{1},g_{2}) = 4\left(\begin{array}{cc}
		\frac{\langle\Delta\hat{X}^{2}\rangle_{i}}{(M+1)^{2}\bar{z}^{2}}+\frac{\langle\Delta\hat{P}^{2}\rangle_{i}}{2k^{2}}+\frac{\mathrm{Cov}_{i}(\hat{X},\hat{P})}{(M+1)\bar{z}k} & \frac{\langle\Delta\hat{X}^{2}\rangle_{i}}{(M+1)^{2}\bar{z}^{2}}+\frac{\mathrm{Cov}_{i}(\hat{X},\hat{P})}{(M+1)\bar{z}k} \\
		\frac{\langle\Delta\hat{X}^{2}\rangle_{i}}{(M+1)^{2}\bar{z}^{2}}+\frac{\mathrm{Cov}_{i}(\hat{X},\hat{P})}{(M+1)\bar{z}k} & \frac{\langle\Delta\hat{X}^{2}\rangle_{i}}{(M+1)^{2}\bar{z}^{2}}+\frac{\langle\Delta\hat{P}^{2}\rangle_{i}}{2k^{2}}+\frac{\mathrm{Cov}_{i}(\hat{X},\hat{P})}{(M+1)\bar{z}k}
	\end{array}\right). \label{eq:S28}
\end{equation}
Combining with the Jacobian matrix in Eq. (\ref{eq:S16}), the corresponding QFIM in the reparameterization $(\bar{\theta},g_{2})^{\mathsf{T}}$ is derived as
\begin{align*}
    \mathcal{Q}^{(\mathrm{BCR})}(\bar{\theta},g_{2}) &= J^{\mathsf{T}}\mathcal{Q}^{(\mathrm{BCR})}(g_{1},g_{2})J\\
    &=4\left(
	\begin{array}{cc}
	    M^{2}\langle\Delta\hat{X}^{2}\rangle_{i}+\frac{(M+1)^{2}M^{2}\bar{z}^{2}\langle\Delta\hat{P}^{2}\rangle_{i}}{2k^{2}}+\frac{(M+1)M^{2}\bar{z}\mathrm{Cov}_{i}(\hat{X},\hat{P})}{k} & -\frac{(M+1)M\bar{z}\langle\Delta\hat{P}^{2}\rangle_{i}}{2k^{2}} \\
	    -\frac{(M+1)M\bar{z}\langle\Delta\hat{P}^{2}\rangle_{i}}{2k^{2}} & \frac{\langle\Delta\hat{P}^{2}\rangle_{i}}{k^{2}}
	\end{array}\right).
\end{align*}
Substituting it into Eq. (\ref{eq:S8}), we can calculate the partial QFI for estimating $\bar{\theta}$ in the presence of a nuisance parameter $g_{2}$ within the classical-label strategy is
\begin{align}
    \mathcal{Q}_{\bar{\theta}|g_{2}}^{(\mathrm{BCR})} &= \mathcal{Q}_{\bar{\theta};\bar{\theta}}^{(\mathrm{BCR})}-\mathcal{Q}_{\bar{\theta};g_{2}}^{(\mathrm{BCR})}\left[\mathcal{Q}_{g_{2};g_{2}}^{(\mathrm{BCR})}\right]^{-1}\mathcal{Q}_{g_{2};\bar{\theta}}^{(\mathrm{BCR})} \nonumber\\
    &= M^{2}(M+1)^{2}\bar{z}^{2}\langle\Delta\hat{P}^{2}\rangle_{i}/k^{2}+4M^{2}(M+1)\bar{z}\mathrm{Cov}_{i}(\hat{X},\hat{P})/k+4M^{2}\langle\Delta\hat{X}^{2}\rangle_{i}, \label{eq:S29}
\end{align}
where the leading $M^{4}$ contribution survives in the partial QFI,implying an asymptotic $1/M^{4}$ scaling of the variance bound and, equivalently, an asymptotic $1/M^{2}$-scaling precision
\begin{equation}
	\lim\limits_{M\to\infty}\mathcal{Q}_{\bar{\theta}|g_{2}}^{(\mathrm{BCR})}/M^{4} = \frac{\bar{z}^{2}\langle\Delta\hat{P}^{2}\rangle_{i}}{k^{2}}. \label{eq:S30}
\end{equation}
Therefore, the classical-label strategy yields the same asymptotic $1/M^{2}$-scaling precision limit with the coherent-control strategy
\begin{equation}
	\lim\limits_{M\to\infty}\delta\bar{\theta}_{\mathrm{QL}}^{(\mathrm{BCR})}/M^{-2} = \lim\limits_{M\to\infty}M^{2}/\sqrt{\nu\mathcal{Q}_{\bar{\theta}|g_{2}}^{(\mathrm{BCR})}} = \frac{k}{\bar{z}\sqrt{\nu\langle\Delta\hat{P}^{2}\rangle_{i}}}. \label{eq:S31}
\end{equation}

\subsection{Estimation from an unlabeled probabilistic route mixture}
In the preceding coherently-control and classically-labeled routing strategies, the estimation can use the joint state of the probe and the route-control degree of freedom. We next consider the case in which the route label is discarded before the final measurement. In this case, the final probe state can be obtained by tracing out the ancilla
\begin{equation}
	\hat{\rho}_{f}^{(\mathrm{mix})} = \mathrm{Tr}_{\mathrm{ancilla}}\left[\hat{\rho}_{f}^{(\mathrm{BCR})}\right] = \frac{1}{2}|\psi_{+}\rangle\langle\psi_{+}|+\frac{1}{2}|\psi_{-}\rangle\langle\psi_{-}|, \label{eq:S32}
\end{equation}
which is a probabilistic mixture of the two pure route-dependent states $|\psi_{+}\rangle$ and $|\psi_{-}\rangle$. Due to the convexity of QFIM, the QFIM for estimating $g_{1}$ and $g_{2}$ from $\hat{\rho}_{f}^{(\mathrm{mix})}$ satisfies
\begin{equation}
	\mathcal{Q}(g_{1},g_{2}|\hat{\rho}_{f}^{(\mathrm{mix})}) \le \frac{1}{2}\mathcal{Q}(g_{1},g_{2}||\psi_{+}\rangle)+\frac{1}{2}\mathcal{Q}(g_{1},g_{2}||\psi_{-}\rangle) = \mathcal{Q}^{(\mathrm{BCR})}(g_{1},g_{2}). \label{eq:S33}
\end{equation}
Although it is intractable to determine whether the above bound is always attainable, we can show that the $M^{4}$-scaling contribution to the partial QFI remains attainable in the perturbative regime of small parameters, i.e., for $g_{1}\ll 1$ and $g_{2}\ll 1$. In this case, it is equivalent to calculate the QFIM at $\bm{g}=\bm{0}$. Here we denote
\begin{equation*}
	\hat{U}_{\bar{z}} = \exp\left[-\mathrm{i}\frac{(M+1)\bar{z}}{2k}\hat{P}^{2}\right], \quad \hat{V}_{+} = \exp\left(-\mathrm{i}\frac{g_{1}}{k}\hat{P}\right)\exp\left[-\mathrm{i}\frac{g_{1}+g_{2}}{(M+1)\bar{z}}\hat{X}\right], \quad \hat{V}_{-} = \exp\left(-\mathrm{i}\frac{g_{2}}{k}\hat{P}\right)\exp\left[-\mathrm{i}\frac{g_{1}+g_{2}}{(M+1)\bar{z}}\hat{X}\right],
\end{equation*}
then we have
\begin{equation*}
	\hat{U}_{+}=\hat{U}_{\bar{z}}\hat{V}_{+}, \quad \hat{U}_{-}=\hat{U}_{\bar{z}}\hat{V}_{-}.
\end{equation*}
The final probe state at $\bm{g}=\bm{0}$ can be written as
\begin{equation}
	\left.\hat{\rho}_{f}^{(\mathrm{mix})}\right|_{\bm{g}=\bm{0}} = \hat{U}_{\bar{z}}|\psi_{i}\rangle\langle\psi_{i}|\hat{U}_{\bar{z}}^{\dagger}, \label{eq:S34}
\end{equation}
which is a pure state. However, variations of parameters at $\bm{g}=\bm{0}$ still drive the final probe state into a mixed state. Therefore, we must directly apply Eq. (\ref{eq:S1}) to calculate the QFIM of $g_{1}$ and $g_{2}$ from $\hat{\rho}_{f}^{(\mathrm{mix})}$ at $\bm{g}=\bm{0}$. Therefore, we first calculate the derivatives of $\hat{\rho}_{f}^{(\mathrm{mix})}$ with respect to $g_{1}$ and $g_{2}$ at $\bm{g}=\bm{0}$, which can be obtained from
\begin{align*}
	\left.\frac{\partial\hat{U}_{+}}{\partial g_{1}}\right|_{\bm{g}=\bm{0}} = \hat{U}_{\bar{z}}\left.\frac{\partial\hat{V}_{+}}{\partial g_{1}}\right|_{\bm{g}=\bm{0}} = -\mathrm{i}\hat{U}_{\bar{z}}\hat{H}_{1}, \quad &\left.\frac{\partial\hat{U}_{+}}{\partial g_{2}}\right|_{\bm{g}=\bm{0}} = \hat{U}_{\bar{z}}\left.\frac{\partial\hat{V}_{+}}{\partial g_{2}}\right|_{\bm{g}=\bm{0}} = -\mathrm{i}\hat{U}_{\bar{z}}\hat{H}_{2}, \\
	\left.\frac{\partial\hat{U}_{-}}{\partial g_{1}}\right|_{\bm{g}=\bm{0}} = \hat{U}_{\bar{z}}\left.\frac{\partial\hat{V}_{-}}{\partial g_{1}}\right|_{\bm{g}=\bm{0}} = -\mathrm{i}\hat{U}_{\bar{z}}\hat{H}_{2}, \quad &\left.\frac{\partial\hat{U}_{-}}{\partial g_{2}}\right|_{\bm{g}=\bm{0}} = \hat{U}_{\bar{z}}\left.\frac{\partial\hat{V}_{-}}{\partial g_{2}}\right|_{\bm{g}=\bm{0}} = -\mathrm{i}\hat{U}_{\bar{z}}\hat{H}_{1},
\end{align*}
where
\begin{equation*}
	\hat{H}_{1} = \frac{\hat{P}}{k}+\frac{\hat{X}}{(M+1)\bar{z}}, \quad \hat{H}_{2} = \frac{\hat{X}}{(M+1)\bar{z}},
\end{equation*}
which are Hermitian operators. Then the derivatives of $\hat{\rho}_{f}^{(\mathrm{mix})}$ are given by
\begin{align*}
	\left.\frac{\partial\hat{\rho}_{f}^{(\mathrm{mix})}}{\partial g_{1}}\right|_{\bm{g}=\bm{0}} &= -\frac{\mathrm{i}}{2}\hat{U}_{\bar{z}}\hat{H}_{1}|\psi_{i}\rangle\langle\psi_{i}|\hat{U}_{\bar{z}}^{\dagger}+\frac{\mathrm{i}}{2}\hat{U}_{\bar{z}}|\psi_{i}\rangle\langle\psi_{i}|\hat{H}_{1}{U}_{\bar{z}}^{\dagger}-\frac{\mathrm{i}}{2}\hat{U}_{\bar{z}}\hat{H}_{2}|\psi_{i}\rangle\langle\psi_{i}|\hat{U}_{\bar{z}}^{\dagger}+\frac{\mathrm{i}}{2}\hat{U}_{\bar{z}}|\psi_{i}\rangle\langle\psi_{i}|\hat{H}_{2}{U}_{\bar{z}}^{\dagger}, \\
	\left.\frac{\partial\hat{\rho}_{f}^{(\mathrm{mix})}}{\partial g_{2}}\right|_{\bm{g}=\bm{0}} &= -\frac{\mathrm{i}}{2}\hat{U}_{\bar{z}}\hat{H}_{2}|\psi_{i}\rangle\langle\psi_{i}|\hat{U}_{\bar{z}}^{\dagger}+\frac{\mathrm{i}}{2}\hat{U}_{\bar{z}}|\psi_{i}\rangle\langle\psi_{i}|\hat{H}_{2}{U}_{\bar{z}}^{\dagger}-\frac{\mathrm{i}}{2}\hat{U}_{\bar{z}}\hat{H}_{1}|\psi_{i}\rangle\langle\psi_{i}|\hat{U}_{\bar{z}}^{\dagger}+\frac{\mathrm{i}}{2}\hat{U}_{\bar{z}}|\psi_{i}\rangle\langle\psi_{i}|\hat{H}_{1}{U}_{\bar{z}}^{\dagger},
\end{align*}
which can be further derived as
\begin{equation}
	\left.\frac{\partial\hat{\rho}_{f}^{(\mathrm{mix})}}{\partial g_{1}}\right|_{\bm{g}=\bm{0}} = \left.\frac{\partial\hat{\rho}_{f}^{(\mathrm{mix})}}{\partial g_{2}}\right|_{\bm{g}=\bm{0}} = -\frac{\mathrm{i}}{2}\hat{U}_{\bar{z}}\left[\hat{H}_{1}+\hat{H}_{2},|\psi_{i}\rangle\langle\psi_{i}|\right]{U}_{\bar{z}}^{\dagger}. \label{eq:S35}
\end{equation}
Here we denote
\begin{equation*}
	\hat{H}_{0} = \hat{H}_{1}+\hat{H}_{2} = \frac{\hat{P}}{k}+\frac{2\hat{X}}{(M+1)\bar{z}},
\end{equation*}
and substitute Eq. (\ref{eq:S34}) and Eq. (\ref{eq:S35}) into Eq. (\ref{eq:S2}), the SLD operators of parameters $g_{1}$ and $g_{2}$ are given by
\begin{equation}
	\left.\hat{L}_{1}\right|_{\bm{g}=\bm{0}} = \left.\hat{L}_{2}\right|_{\bm{g}=\bm{0}} = -\mathrm{i}\hat{U}_{\bar{z}}\left[\hat{H}_{0},|\psi_{i}\rangle\langle\psi_{i}|\right]{U}_{\bar{z}}^{\dagger}. \label{eq:S36}
\end{equation}
Substituting Eq. (\ref{eq:S34}) and Eq. (\ref{eq:S36}) into Eq. (\ref{eq:S1}), we can calculate the corresponding QFIM at $\bm{g}=\bm{0}$ as
\begin{equation*}
	\left.\mathcal{Q}^{(\mathrm{mix})}(g_{1},g_{2})\right|_{\bm{g}=\bm{0}} = \mathrm{Var}_{|\psi_{i}\rangle}(\hat{H}_{0})\left(\begin{array}{cc}
		1 & 1 \\
		1 & 1
	\end{array}\right),
\end{equation*}
which is a singular matrix. Therefore, in the reparameterization $(\bar{\theta},g_{2})^{\mathsf{T}}$, the corresponding QFIM is determined as
\begin{equation*}
	\left.\mathcal{Q}^{(\mathrm{mix})}(\bar{\theta},g_{2})\right|_{\bm{g}=\bm{0}} = \mathrm{Var}_{|\psi_{i}\rangle}(\hat{H}_{0})\left(\begin{array}{cc}
		M^{2}(M+1)^{2}\bar{z}^{2} & 0 \\
		0 & 0
	\end{array}\right),
\end{equation*}
which indicates that only the parameter $\bar{\theta}$ is estimable and the partial QFI is directly given by
\begin{equation}
	\left.\mathcal{Q}_{\bar{\theta}|g_{2}}^{(\mathrm{mix})}\right|_{\bm{g}=\bm{0}} = \left.\mathcal{Q}_{\bar{\theta};\bar{\theta}}^{(\mathrm{mix})}\right|_{\bm{g}=\bm{0}} = M^{2}(M+1)^{2}\bar{z}^{2}\mathrm{Var}_{|\psi_{i}\rangle}(\hat{H}_{0}) = \mathcal{Q}_{\bar{\theta}|g_{2}}^{(\mathrm{BCR})}. \label{eq:S37}
\end{equation}
Thus, in the perturbative regime, discarding the route label does not change the partial QFI relative to classically labeled bidirectional routing. The unlabeled mixture therefore retains the $1/M^{2}$ uncertainty scaling and can outperform either individual route in estimating the global parameter in the presence of nuisance parameters.

\section{Experimental Configurations}
In this section, We provide additional details on the reconfiguration of the sensing network and the compensation of the unwanted polarization phase.

\subsection{Configuration of sensing network}
\begin{figure}[hbt]
	\centering
	\includegraphics[width=\linewidth]{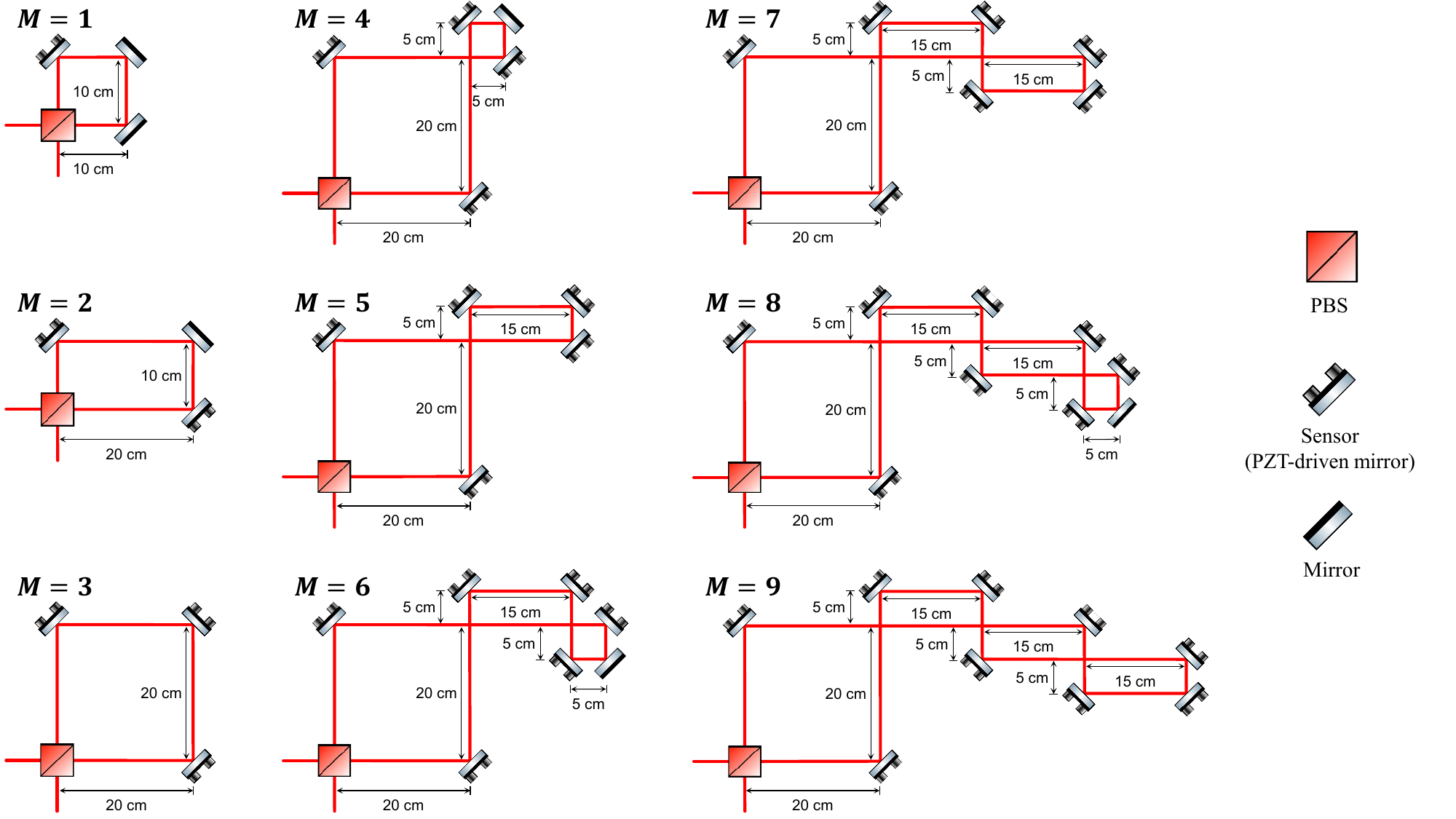}
	\caption{\label{fig:S1} Configuration of sensing network with different number of sensors.}
\end{figure}

\begin{figure}[hbt]
	\centering
	\includegraphics[width=0.6\linewidth]{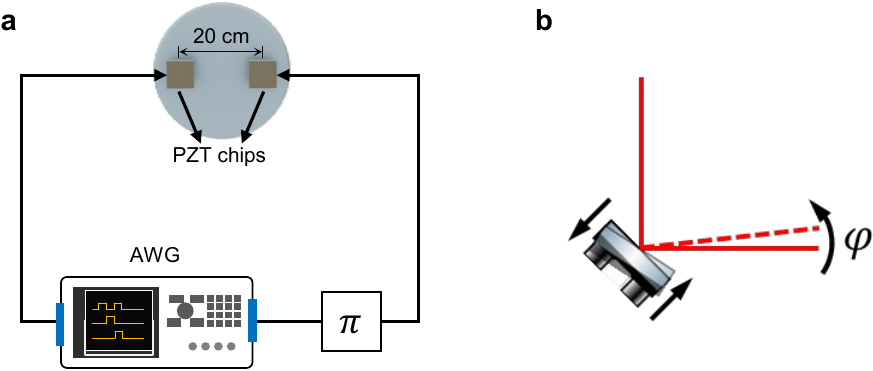}
	\caption{\label{fig:S2} Design of sensing node. \textbf{\textsf{a}} Configuration and applied signals of PZT chips. \textbf{\textsf{b}} Tilt generation.}
\end{figure}

In our experiments, we scale the size of the sensing network by reconfiguring the Sagnac interferometer, as shown in Figure \ref{fig:S1}. To ensure effective weak value amplification, the interferometer is configured with an odd number of mirrors. Accordingly, mirrors driven by PZT chips serve as sensors, whereas mirrors without PZT actuation are used solely to complete the interferometric path. In Figure \ref{fig:S1}, we illustrate the interferometer setup for sensing networks comprising 1 sensor to 9 sensors. The mean distance between adjacent network nodes is fixed at $\bar{z}=\SI{20}{\cm}$, giving a total cyclic path length of $(M+1)\times\SI{20}{\cm}$. The increase in total path length reflects the physical expansion of the sensing network as additional nodes are introduced, conventional distributed architectures likewise require additional transmission links when the number of sensing nodes increases. Additionally, the sensing node is realized using a mirror actuated by two PZT chips in our experiments, and its design is presented in Figure \ref{fig:S2}. Two PZT chips are bonded to the rear surface of the mirror and separated by $\SI{20}{\mm}$ along the horizontal direction. Each PZT actuator produces a displacement of approximately $\SI{22}{\nm}$ under a $\SI{1}{\V}$ drive. As shown in Figure \ref{fig:S2}\textbf{\textsf{a}}, we apply two sinusoidal signals with opposite phases to the two PZT chips. Thus, a $\SI{1}{\V}$ peak-to-peak drive signal corresponds to a $\SI{1.1}{\micro\radian}$ tilt angle on the mirror, which leads to a $\SI{2.2}{\micro\radian}$ beam-tilt modulation.

\subsection{Compensation for relative phase}
In practice, experimental imperfections, such as interferometer misalignment and polarization-dependent mirror reflectivities, can introduce an unwanted relative phase between the horizontal and vertical components in the output polarization state of the probe, thereby degrading the performance of weak value amplification. In the following, we present a compensation method to mitigate this relative phase.

\begin{figure}[hbt]
	\centering
	\includegraphics[width=0.75\linewidth]{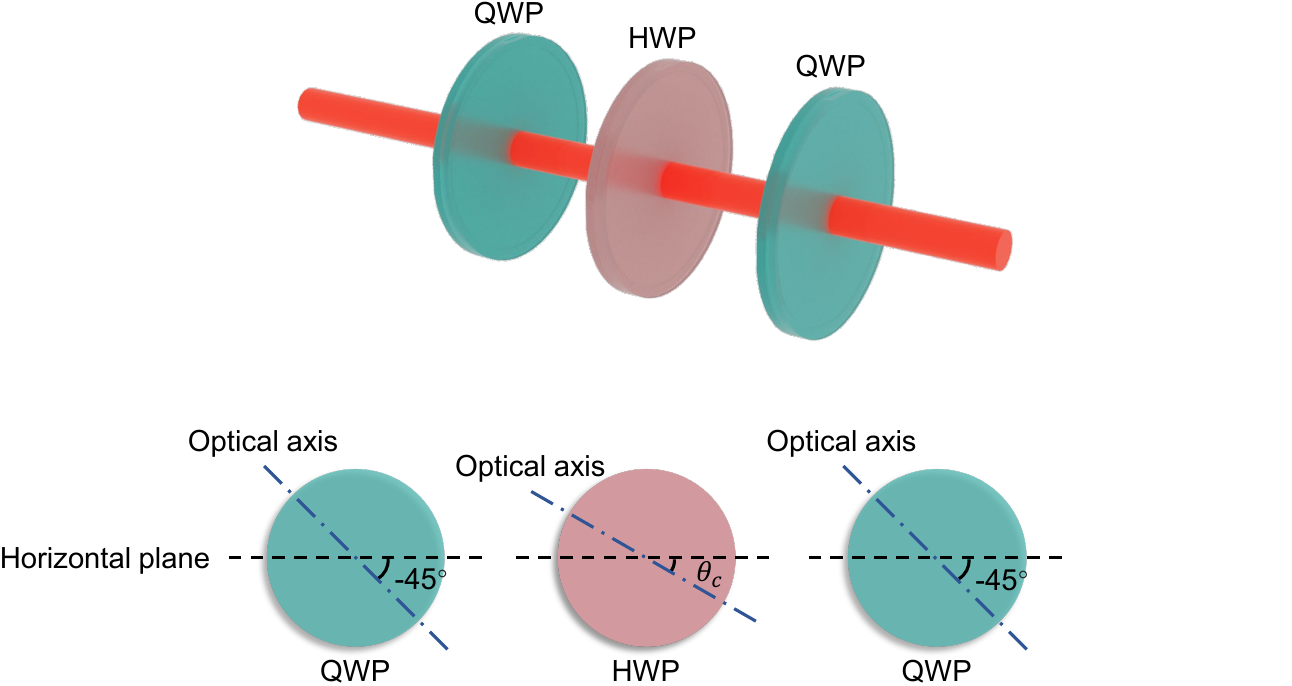}
	\caption{\label{fig:S3} Compensation method for the unwanted relative phase.}
\end{figure}

Up to a global phase, an arbitrary unitary transformation of the polarization qubit can be represented by an element of SU(2), a three-parameter Lie group that is topologically equivalent to the hypersphere $S^{3}$. A convenient parameterization employs Euler angles: any such unitary can be expressed as a product of three rotations, and can be written as
\begin{equation}
	\hat{U}(\varphi,\xi,\zeta)=\hat{R}_{y}({\varphi})\hat{R}_{z}(-\xi)\hat{R}_{y}(\zeta), \label{eq:S023}
\end{equation}
with independent real parameters $(\varphi,\xi,\zeta)$, and $\hat{R}_{y}$ and $\hat{R}_{z}$ denote the rotations about the $y$ and $z$ axes, respectively. Since SU(2) is the double cover of SO(3), these operations correspond to rotations of the Bloch (or Poincaré) sphere. Therefore, rotations $\hat{R}_{y}$ and $\hat{R}_{z}$ are given by
\begin{equation*}
	\hat{R}_{y}(\theta) = \cos(\theta)\hat{\mathbb{I}}-\mathrm{i}\sin(\theta)\hat{\sigma}_{y}, \quad \hat{R}_{z}(\theta) = \cos(\theta)\hat{\mathbb{I}}-\mathrm{i}\sin(\theta)\hat{\sigma}_{z},
\end{equation*}
where
\begin{equation*}
	\hat{\sigma}_{y} = \left(
	\begin{array}{cc}
		0 & -\mathrm{i} \\
		\mathrm{i} & 0
	\end{array}
	\right), \qquad
	\hat{\sigma}_{z} = \left(
	\begin{array}{cc}
		1 & 0 \\
		0 & -1
	\end{array}
	\right),
\end{equation*}
are the Pauli-Y and Pauli-Z matrices, respectively.

Experimentally, an arbitrary SU(2) transformation on the polarization state can be realized with the standard QWP–HWP–QWP sequence
\begin{equation*}
	\hat{U} = \mathrm{QWP}(\eta_{1})\,\mathrm{HWP}(\tau)\,\mathrm{QWP}(\eta_{2}),
\end{equation*}
where the arguments denote the respective angles of the plates’ fast axes with respect to the horizontal reference, as illustrated in Figure \ref{fig:S3}.

According to the Jones matrix representation, the operations $\mathrm{QWP}(\eta)$ and $\mathrm{HWP}(\tau)$ are given by
\begin{equation*}
	\mathrm{QWP}(\eta) = \hat{R}_{y}(\eta)\hat{Q}_{0}\hat{R}_{y}(-\eta), \quad \mathrm{HWP}(\tau) = \hat{R}_{y}(\tau)\hat{H}_{0}\hat{R}_{y}(-\tau),
\end{equation*}
where
\begin{equation*}
	\hat{Q}_{0} = \left(
	\begin{array}{cc}
		1 & 0 \\
		0 & \mathrm{i}
	\end{array}
	\right), \qquad
	\hat{H}_{0} = \left(
	\begin{array}{cc}
		1 & 0 \\
		0 & -1
	\end{array}
	\right).
\end{equation*}
Then the angles of these three wave plates can be obtained as
\begin{equation*}
	\eta_{1} = \varphi-\frac{\pi}{4}, \quad \eta_{2} = -\zeta-\frac{\pi}{4}, \quad \tau = \frac{1}{2}(\varphi+\xi-\zeta)-\frac{\pi}{4}.
\end{equation*}

To compensate the unwanted relative phase $\delta\theta$ introduced by imperfections in the Sagnac interferometer, we implement the compensation operation
\begin{equation*}
	\hat{U}_{c} = \hat{R}_{z}(-\frac{\delta\theta}{2}),
\end{equation*}
which corresponds to Euler-rotation angles
\begin{equation*}
	\varphi=\zeta=0, \quad \xi=\frac{\delta\theta}{2}.
\end{equation*}
Accordingly, in the QWP–HWP–QWP configuration, both QWPs are fixed at $-\ang{45}$, and the HWP is set to $\delta\theta/4-\ang{45}$, thereby canceling the unwanted phase offset.

\section{Supplementary Experimental Results}
\begin{figure}[hbt]
	\centering
	\includegraphics[width=\linewidth]{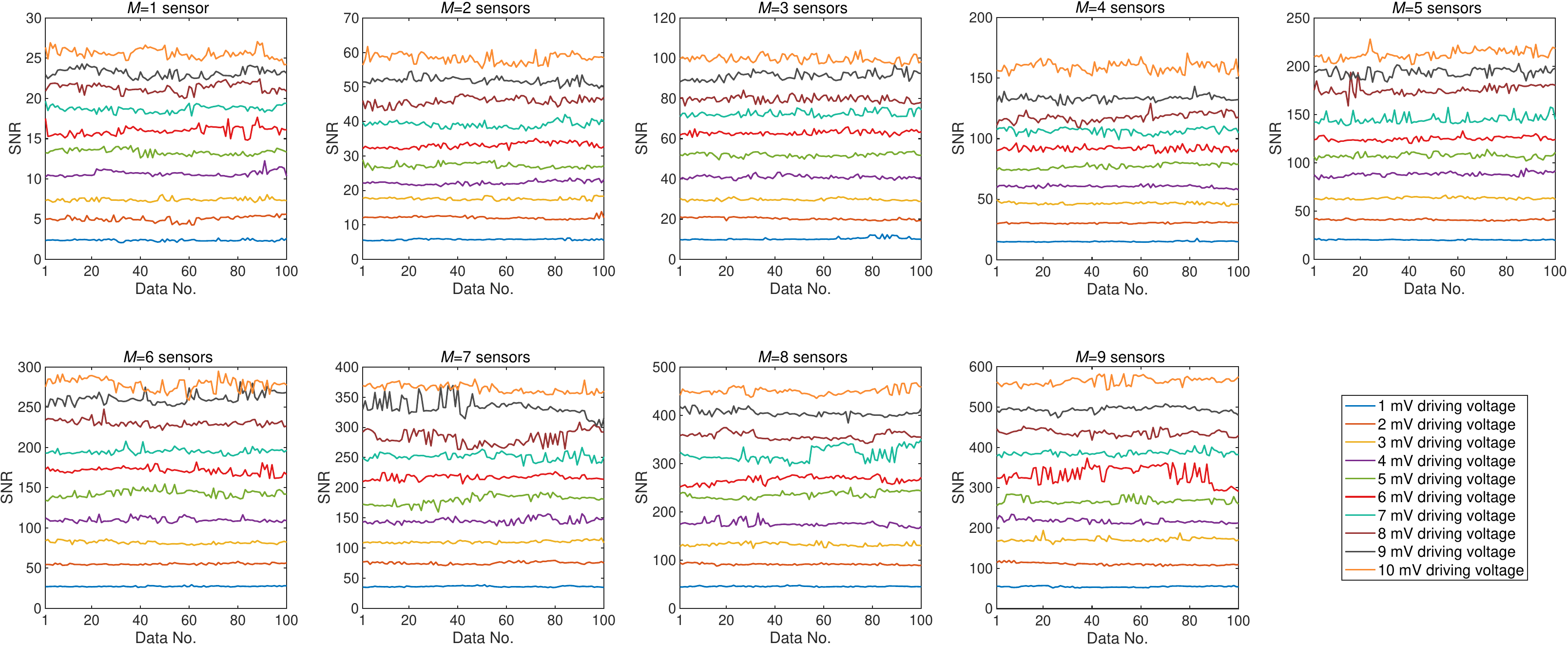}
	\caption{\label{fig:S4} Experimental results of detected SNR.
		Each subfigure shows results for a different number of sensors in the network. Within each subfigure, the solid curves, ordered from bottom to top, correspond to input signal voltages from $\SI{1}{\mV}$ to $\SI{10}{\mV}$, respectively, applied at the sensors. Each curve consists of 100 measured data points.}
\end{figure}

\begin{table*}[hbt]
	\begin{center}
		\begin{minipage}{\textwidth}
			\caption{Experimental results of detected precision.}
			\label{tab:S1}
			\begin{tabular*}{\textwidth}{@{\extracolsep{\fill}}cccccccccc@{\extracolsep{\fill}}}
				\toprule
				Number of sensors & 1 & 2 & 3 & 4 & 5 & 6 & 7 & 8 & 9 \\
				\midrule
				$V_{\min}^{(\mathrm{PZT})}$\footnotemark[1] & $\SI{382.6}{\uV}$ & $\SI{175.3}{\uV}$ & $\SI{98.6}{\uV}$ & $\SI{65.5}{\uV}$ & $\SI{46.8}{\uV}$ & $\SI{35.5}{\uV}$ & $\SI{27.6}{\uV}$ & $\SI{22.2}{\uV}$ & $\SI{18.1}{\uV}$ \\
				$\delta\bar{\varphi}_{\min}^{(\mathrm{det})}$\footnotemark[2] & $\SI{841.8}{\pico\radian}$ & $\SI{385.7}{\pico\radian}$ & $\SI{217.0}{\pico\radian}$ & $\SI{144.1}{\pico\radian}$ & $\SI{103.1}{\pico\radian}$ & $\SI{77.7}{\pico\radian}$ & $\SI{60.6}{\pico\radian}$ & $\SI{48.9}{\pico\radian}$ & $\SI{39.8}{\pico\radian}$ \\
				\bottomrule
			\end{tabular*}
			\footnotetext[1]{Minimal peak-to-peak voltages of signals applied to sensors, corresponding to detected $\mathrm{SNR}=1$ at the spectrum analyzer.}
			\footnotetext[2]{Minimal detectable average tilt angles in experiments.}
		\end{minipage}
	\end{center}
\end{table*}

In our experiments, we synchronously applied sinusoidal drive signals to PZT chips of sensors, and and recorded the peak SNR using the spectrum analyzer. The peak-to-peak voltages of applied signals were varied from $\SI{1}{\mV}$ to $\SI{10}{\mV}$ in $\SI{1}{\mV}$ increments. For each configuration of the sensing network (number of sensors) and applied signal voltage, the experiment was repeated 100 times, and the peak SNR at $\SI{10}{\kHz}$ was recorded from the spectrum analyzer. The detailed results are presented in Figure \ref{fig:S4}.

For each sensing-network configuration (i.e., for a fixed number of sensors), the SNR scales linearly with the applied signal voltage, and this linear relation can be determined from the experimental data presented above. Imposing $\mathrm{SNR}=1$ defines the minimum applicable voltage $V_{\min}^{(\mathrm{PZT})}$ for the PZT chips of sensors, which corresponds to the experimentally determined minimum detectable average tilt angle, $\delta\bar{\varphi}_{\min}^{(\mathrm{det})}$. The detailed results are listed in Table \ref{tab:S1}.
